\documentclass[pdftex,twocolumn,epjc3]{svjour3}          % twocolumn

\RequirePackage[T1]{fontenc}

\smartqed  % flush right qed marks, e.g. at end of proof

\RequirePackage{graphicx}
\RequirePackage{mathptmx}      % use Times fonts if available on your TeX system
\RequirePackage{flushend}
\RequirePackage[numbers,sort&compress]{natbib}
\RequirePackage[colorlinks,citecolor=blue,urlcolor=blue,linkcolor=blue]{hyperref}

\usepackage{graphics}
\usepackage{ucs}
\usepackage[utf8x]{inputenc}
\usepackage[T1]{fontenc}
\usepackage{hyperref}
\usepackage{amsmath,amssymb,amstext}
\usepackage[makeroom]{cancel}
\usepackage{graphicx}
\usepackage{listings}
\usepackage{alg}
\usepackage{breqn}

\journalname{Eur. Phys. J. C}

\begin{document}

\title{Neutron Stars in Palatini $R+\alpha R^2$ and $R+\alpha R^2+\beta Q$ Theories}

\author{Georg Herzog\thanksref{e1,addr1} \and H\`elios Sanchis-Alepuz\thanksref{e2,addr2,addr3}
}

\thankstext{e1}{e-mail: g.herzog@campus.unimib.it}
\thankstext{e2}{e-mail: hsanchisalepuz@gmail.com}

\institute{Dipartimento di Fisica 'G. Occhialini', Universit\`a degli Studi di Milano-Bicocca, Piazza della Scienza 3, I-20126 Milano, Italy\label{addr1} \and Institute of Physics, University of Graz, NAWI Graz, Universit\"atsplatz 5, 8010 Graz, Austria\label{addr2}\and Silicon Austria Labs GmbH, Inffeldgasse 25F, 8010 Graz, Austria\label{addr3}
}

\date{Received: date / Accepted: date}

\maketitle

\begin{abstract}
We study solutions of the stellar structure equations for spherically symmetric objects in modified theories of gravity, where the Einstein-Hilbert Lagrangian is replaced by $f(R)=R+\alpha R^2$ and $f(R,Q)=R+\alpha R^2+\beta Q$, with $R$ being the Ricci scalar curvature, $Q=R_{\mu\nu}R^{\mu\nu}$ and $R_{\mu\nu}$ the Ricci tensor. We work in the Palatini formalism, where the metric and the connection are assumed to be independent dynamical variables. We focus on stellar solutions in the mass-radius region associated to neutron stars. We illustrate the potential impact of the $R^2$ and $Q$ terms by studying a range of viable values of $\alpha$ and $\beta$. Similarly, we use different equations of state (SLy, FPS, HS(DD2) and HS(TMA)) as a simple way to account for the equation of state uncertainty. Our results show that for certain combinations of the $\alpha$  and $\beta$ parameters and equation of state, the effect of modifications of general relativity on the properties of stars is sizeable.  Therefore, with increasing accuracy in the determination of the equation of state for neutron stars, astrophysical observations may serve as discriminators of modifications of General Relativity.
\end{abstract}

\section{Introduction}
General Relativity (GR) as the simplest realisation of a geometric description of gravity has so far shown, perhaps unexpectedly, total agreement with astrophysical and cosmological observations \cite{Will:2014kxa,Berti:2015itd}, as long as the general picture is accepted that the dark sectors, as needed for the $\Lambda$CDM model, are related to the particle and energy content of the Universe.
Despite this tremendous success, there are reasons to investigate  modifications of GR. First, GR breaks down in the high curvature regime. Second, it is only a classical field theory without any quantum effects and which is not (perturbatively) renormalisable as a quantum field theory unless quadratic curvature corrections are added to the Lagrangian \cite{Stelle:1976gc}. Moreover, it is conceivable that all or part of the effects currently attributed to dark sectors are in reality a manifestation of different gravitational dynamics \cite{Clifton:2011jh,deMartino:2015zsa,Ferreira:2019xrr,Frusciante:2019xia}.

Some of the (theoretical) problems of GR can be ameliorated in modified gravitational theories, like $f(R)$ and $f(R,Q)$ theories in which the Einstein-Hilbert Lagrangian is modified by adding polynomial terms in the Ricci scalar $R$ or in the contraction of the Ricci curvature $Q=R_{\mu\nu}R^{\mu\nu}$ (see e.g. \cite{Sotiriou:2008rp,DeFelice:2010aj,Capozziello:2010zz,Olmo:2011uz,Capozziello:2015lza,Nojiri:2017ncd} and references therein). From a different point of view, if the different terms in a gravity action are considered as quantum operators in the spirit of effective quantum field theories, studying their renormalisation group evolution indicates that $f(R)$ and $f(R,Q)$ terms render the theory asymptotically safe (and hence viable as a quantum theory) in the Planckian and super-Planckian regime \cite{Codello:2007bd,Hindmarsh:2012rc,Falls:2013bv,Falls:2017lst,Alkofer:2018fxj}. Finally, classical theories of modified  gravity can also serve as effective descriptions of their more fundamental quantum counterparts (see e.g. \cite{Olmo:2008nf,Bonanno:2012jy}).

Studying the compatibility of modified theories of gravity with phenomenology is a highly non-trivial issue since an interpretation of observations involves a combination of many different and uncertain physical mechanisms. As mentioned above, cosmological observations compatible with GR and dark sectors can be also made compatible with certain modifications of gravity and different contributions from dark sectors \cite{Pinto:2018rfg}. Inflationary scenarios can also be expressed in terms of modified theories of gravity (see e.g. \cite{Antoniadis:2018ywb,Antoniadis:2018yfq,Tenkanen:2020dge,Gialamas:2019nly,Gialamas:2020snr}). It has been suggested that modifications of GR can be identified via the detection of new modes in gravitational waves \cite{Capozziello:2008rq,Alves:2009eg,Corda:2010zza,DeLaurentis:2011tp,Rizwana:2016qdq,Liang:2017ahj,Capozziello:2019klx,Katsuragawa:2019uto}. Another possibility to investigate deviations from GR is with astrophysical observations. Different gravitational dynamics reflects into different stellar equations of structure \cite{Kainulainen:2006wz,Olmo:2012er,Berti:2015itd,Paschalidis:2016vmz,Olmo:2019flu,Astashenok:2020isy} and hence, in particular, in mass to radius ratios of neutron stars different to those expected from GR (see e.g. \cite{Arapoglu:2010rz,Deliduman:2011nw,Astashenok:2013vza,Astashenok:2014nua,Astashenok:2014dja,Capozziello:2015yza,Arapoglu:2016ozr,Astashenok:2017dpo,Pannia:2016qbj,Astashenok:2020qds,Astashenok:2021peo,Odintsov:2021qbq,Odintsov:2021nqa}). This is particularly interesting with the growing volume of observational data on neutron stars and the discovery of compact objects that are difficult to accomodate in a standard GR picture (see e.g. \cite{TheLIGOScientific:2017qsa,Alsing:2017bbc,Abbott:2018wiz,Tsokaros:2020hli}). These measurements are, unfortunately, riddled with enormous uncertainties coming from the equation of state (EoS) describing the interior of neutron stars and can as well be interpreted as providing a handle on EoS assuming the theory of gravity is GR \cite{Abbott:2018exr}. However, as we will show in this paper, the changes in the mass-radius relation for neutron stars coming from the uncertainty in the EoS are sometimes comparable to the differences generated by the modifications of the gravity theory, which implies that if, in the future, reliable ab-initio calculations of the EoS were available, astrophysical observations could be used as discriminators for gravity theories.

In this paper we shall investigate the influence of two particular models of $f(R)$ and $f(R,Q)$ theories, namely $f(R)=R+\alpha R^2$ and $f(R,Q)=R+\alpha R^2 + \beta Q$ on the mass-radius relation for neutron stars. We study those theories in the Palatini formalism in which the dynamical degrees of freedom are the metric and the affine connection \cite{Olmo:2011uz}. We will perform our calculations using a small number of representative EoS with the goal of estimating the uncertainty stemming from the EoS and from the theory of gravity (in our case simply encoded in the values of $\alpha$ and $\beta$).

The paper is organised as follows. We sketch the derivation of the equations of stellar structure for Palatini theories in Sec.~\ref{sec_stellarstructureeqs} and discuss several aspects of the EoS chosen in Sec.~\ref{sec_EOS}. Our results are shown in Sec.~\ref{sec_results} and we conclude a discussion of the implications of our results on the validity of theories studied herein and of possible future work.

\section{Stellar Structure Equations for $f(R,Q)$ Palatini Theories}\label{sec_stellarstructureeqs}

We discuss here the main aspects of the equations of stellar structure solved in this work. We consider only the simplified case of non-rotating and static stars, whose only observables are their mass and radius. The equations for $f(R,Q)$ theories were derived in \cite{Olmo:2012er} and we refer to that paper for further details (see also \cite{Kainulainen:2006wz,Pannia:2016qbj}). In GR, the equivalent equations are the well-known Tolman-Oppenheimer-Volkoff (TOV) equations. 

In Palatini $f(R,Q)$ theories the action is given by
\begin{equation}\label{eq_action}
S[g,\Gamma,\psi_m]=\frac{1}{2\kappa}\int d^4x \sqrt{-g}f(R,Q) +S_m[g,\psi_m],
\end{equation}
where $R$ is the Ricci scalar, $Q=R_{\mu\nu}R^{\mu\nu}$ the aforementioned contraction of two Ricci tensors, with $R_{\mu \nu}=R_{\mu \rho \nu}^{\rho}$ and the Riemann tensor given by $R^d_{abc}\equiv \partial_b\Gamma^d_{ac}-\partial_c\Gamma^d_{ab}+\Gamma^e_{ac}\Gamma^d_{eb}-\Gamma^e_{ab}\Gamma^d_{ec}$ with $\Gamma$ the connection coefficients, $g_{\mu\nu}$ is the metric, $\kappa\equiv 8\pi G$ and $S_m[g,\psi_m]$ is the matter action. In the Palatini formalism the equations of motion are obtained after varying the action with respect to both, the metric and the connection
\begin{eqnarray}
f_R R_{\mu\nu}-\frac{f}{2}g_{\mu\nu}+2f_QR_{\mu\alpha}{R^\alpha}_\nu &=& \kappa T_{\mu\nu}\label{metrEqfRQ}\\
\nabla_{\beta}\left[\sqrt{-g}\left(f_R g^{\mu\nu}+2f_Q R^{\mu\nu}\right)\right]&=&0\label{connEqfRQ}  \ ,
\end{eqnarray}
with $f_R \equiv \frac{\partial f}{\partial R}$ and $f_Q \equiv \frac{\partial f}{\partial Q}$.
%and $T_{\mu\nu}$ the stress-energy tensor of a perfect fluid which is given by
%\begin{equation}\label{eq_StressEnergyTensorPerfectFluid}
%T^{\mu \nu}=\left(\rho+\frac{P}{c^2}\right)u^{\mu}u^{\nu}-Pg^{\mu \nu}
%\end{equation}
%where $\rho$ is the energy density and $P$ the pressure.
One can introduce an auxiliary metric $h_{\mu\nu}$ via the (matter-content dependent) mapping $\sqrt{-g}\left(f_R g^{\mu\nu}+2f_Q R^{\mu\nu}\right)\equiv\sqrt{-h}h^{\mu\nu}$ such that the equation for the connection becomes $\nabla_\beta \left[\sqrt{-h}h^{\mu\nu}\right]=0$ and the connection $\Gamma$ can thus be written as the Levi-Civita connection for $h_{\mu\nu}$. Therefore, the Ricci tensor in Eqs.~\ref{metrEqfRQ} and \ref{connEqfRQ} can be written as the \textit{standard} metric Ricci tensor of the metric $h$ which we denote as $R_{\mu\nu}(h)$.
Introducing ${\Sigma_\alpha} ^\nu := \left(f_R \delta_\alpha^\nu +2f_Q {B_\alpha}^\nu\right)$, with ${B_\alpha}^\nu$ given as $ {B_\alpha}^\nu=R_{\alpha\beta}(h)g^{\beta\nu}$, the relation between $g_{\mu\nu}$ and $h_{\mu\nu}$ is \cite{Olmo:2012er}
\begin{equation} \label{eq:h-g}
h_{\mu\nu}=\sqrt{\det\Sigma} {[\Sigma^{-1}]_\mu}^\alpha g_{\alpha\nu}  \ \text{   } , \text{   } \ h^{\mu\nu}=\frac{g^{\mu\alpha}{{\Sigma_\alpha} ^\nu}}{\sqrt{\det\Sigma}} \ .
\end{equation}
The Ricci tensor is given by
\begin{equation}
\label{Rh}
R_{\mu}^{\nu}(h)=R_{\mu\alpha}(h)h^{\alpha\nu}=\frac{1}{\sqrt{\mathrm{det}\Sigma}}\left(\frac{f}{2}\delta_{\mu}^{\nu}+\kappa T_{\mu}^{\nu}\right).
\end{equation}
Note here that, when calculating $R$ and $Q$ from $R(h)$ in $f(R,Q)$, the Ricci tensor $R_{\mu\nu}(h)$ must be contracted with the physical metric $g^{\mu\nu}$, that is, via the matrix $B_{\alpha}^{\nu}$ above \cite{Olmo:2012er}
\begin{equation}
R=B_{\alpha}^{\alpha} \quad \mathrm{and} \quad Q=B_{\alpha}^{\nu}B_{\nu}^{\alpha}~.
\end{equation}

In this paper we will only consider energy-momentum tensors of the form
\begin{dmath}
T_{\mu\nu}=\left(\rho + P\right)u_\mu u_\nu + P g_{\mu\nu}~, 
\end{dmath}
and Lagrangians of the form $f(R)=R+\alpha R^2$ and $f(R,Q)=R+\alpha R^2 + \beta Q$. In these cases we can write the scalars $R$ and $Q$ in terms of $\rho$ and $P$ as $R=-\kappa T$, as in GR, and \cite{Olmo:2012er}
\begin{dmath}
\label{betaQ}
%\scriptstyle
\beta Q=-\left(\tilde f+\frac{\tilde f_R^2}{4f_Q}+2\kappa^2P\right)+\frac{{f_Q}}{16}\left[3\left(R+\frac{\tilde f_R}{f_Q}\right)\pm\sqrt{\left(R+\frac{\tilde f_R}{f_Q}\right)^2-\frac{4\kappa^2(\rho+P)}{f_Q}}\right]^2~,
\end{dmath}
where $\tilde{f}=f(R)=R+\alpha R^2$, $\tilde{f_R}=1+2\alpha R$ and $f_Q=\beta$. The sign in front of the square root must be chosen so as to obtain the right limit at low curvatures.

The stellar structure equations are the dynamical equations for the star's interior metric, parametrised as
\begin{equation}
g_{\mu \nu}=\mathrm{diag}\left(-A(r)e^{2\psi(r)}, \frac{1}{A(r)}, r^2, r^2\mathrm{sin}^2\theta\right)~,
\end{equation}
where $A(r)$ will be written as $A(r)=1-\frac{2M(r)}{r}$. After a tedious but straightforward derivation (see \cite{Olmo:2012er} for details) one arrives at the stellar structure equations for Palatini in the case of $f(R,Q)=R+\alpha R^2 + \beta Q$, which read
\begin{dmath}
\label{dpsifRQ}
\left(\frac{\Omega_r}{\Omega}+\frac{2}{r}\right)\psi_r=\frac{1}{A}\left(\tau_r^r-\frac{\Omega}{S}\tau_t^t\right)-\frac{1}{2}\frac{\Omega_r}{\Omega}\left(2\frac{\Omega_r}{\Omega}+\frac{S_r}{S}\right)-\frac{1}{r}\left(\frac{S_r}{S}-\frac{\Omega_r}{\Omega}\right)+\frac{\Omega_{rr}}{\Omega}~,
\end{dmath}
\begin{dmath}
\label{dmfRQ}
\left(\frac{\Omega_r}{\Omega}+\frac{2}{r}\right)\frac{M_r}{r}=\frac{3\tau_r^r-\frac{\Omega}{S}\tau_t^t}{2}+A\left(\frac{\Omega_{rr}}{\Omega}+\frac{\Omega_r}{\Omega}\left(\frac{2r-3M}{r(r-2M)}-\frac{3}{4}\frac{\Omega_r}{\Omega}\right)\right)~,
\end{dmath}
\begin{dmath}
\label{dPfRQ}
P_r=-\frac{P^{(0)}_r}{\left[1-\alpha(r)\right]}\frac{2}{\left(1\pm\sqrt{1-\beta(r)P^{(0)}_r}\right)}~,
\end{dmath}
where, as before, a subscript denotes partial derivation, i.e. $M_r=\frac{\partial M}{\partial r}$ and $\Omega_{rr}=\frac{\partial^2 \Omega}{\partial r^2}$, and we introduced for compactness
\begin{dmath}
\alpha(r) = \frac{(\rho+P)}{2}\left(\frac{\Omega_P}{\Omega}+\frac{S_P}{S}\right)~,
\end{dmath}
\begin{dmath}
\beta(r) = (2r)\frac{\Omega_P}{\Omega}\left[1-\frac{(\rho+P)}{2}\left(\frac{3}{2}\frac{\Omega_P}{\Omega}-\left\{\frac{\Omega_P}{\Omega}-\frac{S_P}{S}\right\}\right)\right]~,
\end{dmath}
\begin{dmath}
P^{(0)}_r = \frac{(\rho+P)}{r(r-2M)}\left[M-\left(\tau_r^r+\frac{\Omega}{S}\tau_t^t\right)\frac{r^3}{4}\right]~.
\end{dmath}
Here, $\Omega$ and $S$ are given by the relation between the auxiliary metric $h_{\mu \nu}$ and the physical metric $g_{\mu \nu}$
\begin{equation}
h_{tt}=\frac{\sigma_2^2}{\sqrt{\sigma_1\sigma_2}}g_{tt}\equiv Sg_{tt} \quad \mathrm{and} \quad h_{ij}=\sqrt{\sigma_1\sigma_2}g_{ij}\equiv \Omega g_{ij}~,
\end{equation}
where $\sigma_1$ and $\sigma_2$ appear upon rewriting the matrix $\Sigma$ as $\Sigma_{\alpha}^{\nu}=\text{diag(}\sigma_1,\sigma_2,\sigma_2,\sigma_2\text{)}$ from Eq.~\ref{Rh} and are given by \cite{Olmo:2012er}
\begin{equation}
\sigma_1 = \frac{f_R}{2}\pm \sqrt{2f_Q}\sqrt{\lambda^2-\kappa(\rho+P)},
\end{equation}
\begin{equation}
\sigma_2 = \frac{f_R}{2}+\sqrt{2f_Q}\lambda~,
\end{equation}
where
\begin{dmath}
\lambda = \frac{\sqrt{2f_Q}}{8}\left( 3\left( R+\frac{f_R}{f_Q} \right) \pm\sqrt{\left(R+\frac{f_R}{f_Q}\right)^2-\frac{4\kappa(\rho+P)}{f_Q}}\right)~.\nonumber\\
\end{dmath}
In order to recover the GR limit one has to use the positive sign in front of the square root in Eq.~\ref{dPfRQ} and for $\sigma_1$ but the negative sign in $\lambda$ \cite{Olmo:2012er}. Finally, $\tau$ refers to the right-hand side of Eq.~\ref{Rh}, namely $R_{\mu}^{\nu}\equiv\tau_{\mu}^{\nu}$.

Note that, in Eq.~\ref{dmfRQ}, we have $\Omega_{rr}=\Omega_{PP}P_r^2+\Omega_PP_{rr}$ which involves the second derivative $P_{rr}$, in contrast to the standard TOV equations. As it turns out, it is possible to write $P_{rr}$ in terms of the first order derivative $P_r$. The result is \cite{Olmo:2012er} 
\begin{dmath}
\label{PrrPr}
%\scriptstyle
\frac{P_{rr}}{P_{r}}=\frac{P^{(0)}_{rr}}{P^{(0)}_{r}}\left[1+\frac{s}{2}\frac{\beta P^{(0)}_{r}}{\sqrt{1-\beta P^{(0)}_{r}}(1\pm \sqrt{1-\beta P^{(0)}_{r}})}\right]+\left[\frac{\alpha_r}{1-\alpha}+\frac{\pm 1}{2}\frac{\beta_r P^{(0)}_{r}}{\sqrt{1-\beta P^{(0)}_{r}}(1\pm \sqrt{1-\beta P^{(0)}_{r}})}\right]~,
\end{dmath}
with $\alpha_r=\alpha_PP_r$ and $\beta_r=\beta_PP_r$ and 
\begin{dmath}
\label{Prr0Pr0}
%\scriptstyle
\frac{P^{(0)}_{rr}}{P^{(0)}_{r}}=\left[\left(\frac{1+\rho_P}{\rho+P}\right)-\frac{\Phi_P \frac{r^3}{4}}{\left(M-\Phi\frac{r^3}{4}\right)}\right]P_r-\left(\frac{2(r-M)}{r(r-2M)}+\frac{ \frac{3\Phi r^2}{4}}{M-\Phi\frac{r^3}{4} }\right)+M_r\left(\frac{2}{r-2M}+\frac{1}{M-\Phi\frac{r^3}{4}}\right).
\end{dmath}
where we introduced $\Phi\equiv (\tau_r^r+\frac{\Omega}{S}\tau_t^t)$.
In this way, Eqs.~\ref{dpsifRQ}--\ref{dPfRQ} are a closed (after fixing an equation of state) system of equations expressed in terms of $r$, $\rho(P)$, $P$, $M$ and their first radial derivatives only.

\section{Equations of State}\label{sec_EOS}
The essential ingredient to solve the equations of stellar structure is an equation of state that relates the pressure to the energy density inside the star. 

In oder to estimate the different effects on the mass and radius of neutron stars coming from the modifications of the gravity lagrangian and from the uncertainty in the EoS, we use a number of different EoS in analytic as well as tabulated form. Specifically, we use two different tabulated and two different analytic EoS. The analytic equations of state we used are the so-called 
%PLY \cite{Pannia:2016qbj}, 
SLy and FPS \cite{Haensel:2004nu}
%, where PLY is only a polytropic toy model
%\begin{equation}
%\zeta = 2\xi+5.29355,
%\end{equation}
%where $\xi = \text{log}(\rho/\text{g cm}^{-3})$ and $\zeta = %\text{log}(P/\text{dyn cm}^{-2})$. 
The SLy and FPS EoS are analytic parametrisations of the results of many-body calculations with unified effective nuclear Hamiltonians \cite{Friedman:1981qw,Douchin:2001sv}, describing all regions of the neutron star interior from crust to core including its transitions. The analytic paramerisations take care that all thermodynamic conditions involving derivatives of the EoS are fulfilled, an aspect that will become problematic when using tabular data directly, as we will see. 
The SLy and FPS EoS are both parametrised by the following function 
\begin{eqnarray}
  \zeta &=& \frac{a_1+a_2\xi+a_3\xi^3}{1+a_4\,\xi}\,f_0(a_5(\xi-a_6))
\nonumber\\&&
     + (a_7+a_8\xi)\,f_0(a_9(a_{10}-\xi))
\nonumber\\&&
     + (a_{11}+a_{12}\xi)\,f_0(a_{13}(a_{14}-\xi))
\nonumber\\&&
     + (a_{15}+a_{16}\xi)\,f_0(a_{17}(a_{18}-\xi)) ~,
\label{eq_EoS_fit}
\end{eqnarray}
where $\xi = \text{log}(\rho/\text{g cm}^{-3})$ and $\zeta = \text{log}(P/\text{dyn cm}^{-2})$ and $f_0(x)$ is given by \cite{Haensel:2004nu}
\begin{equation}
f_0(x)=\frac{1}{e^x+1}.
\end{equation}
The fitted parameters $a_i$ in Eq.~\ref{eq_EoS_fit} are shown in Table \ref{tab_SLYFPSparams}.

\begin{table}[!t]
\centering
\caption[]{SLy and FPS parameters for Eq. (\protect\ref{eq_EoS_fit}), given in \cite{Haensel:2004nu}.}
\label{tab_SLYFPSparams}
\begin{tabular}{rll|rll}
\hline\hline\rule[-1.4ex]{0pt}{4.3ex}
i & $a_i$(FPS) & $a_i$(SLy) & i & $a_i$(FPS) & $a_i$(SLy)  \\
\hline\rule{0pt}{2.7ex}
$1$    & {6.22} &{6.22}        & ${10}$ &~\, 11.8421   &~\, 11.4950\\
$2$    & {6.121} & {6.121}     & ${11}$ &   $-22.003$  & $-22.775$ \\
$3$    & 0.006004 & 0.005925   & ${12}$ &~\, 1.5552   &~\, 1.5707 \\
$4$    & 0.16345   & 0.16326   & ${13}$ &~\, 9.3      &~\, 4.3    \\
$5$    & 6.50      & 6.48      & ${14}$ &~\, 14.19    &~\, 14.08  \\
$6$    & 11.8440   & 11.4971   & ${15}$ &~\, 23.73    &~\, 27.80  \\
$7$    & 17.24     & 19.105    & ${16}$ &  $-1.508$   & $-1.653$  \\
$8$    & 1.065     & 0.8938    & ${17}$ &~\, 1.79     &~\, 1.50   \\
$9$    & 6.54      & 6.54      & ${18}$ &~\, 15.13    &~\, 14.67 \rule[-1.4ex]{0pt}{0pt}\\
\hline\hline
\end{tabular}
\end{table}

\begin{figure}[htbp]
\centering
  \includegraphics[width=\linewidth]{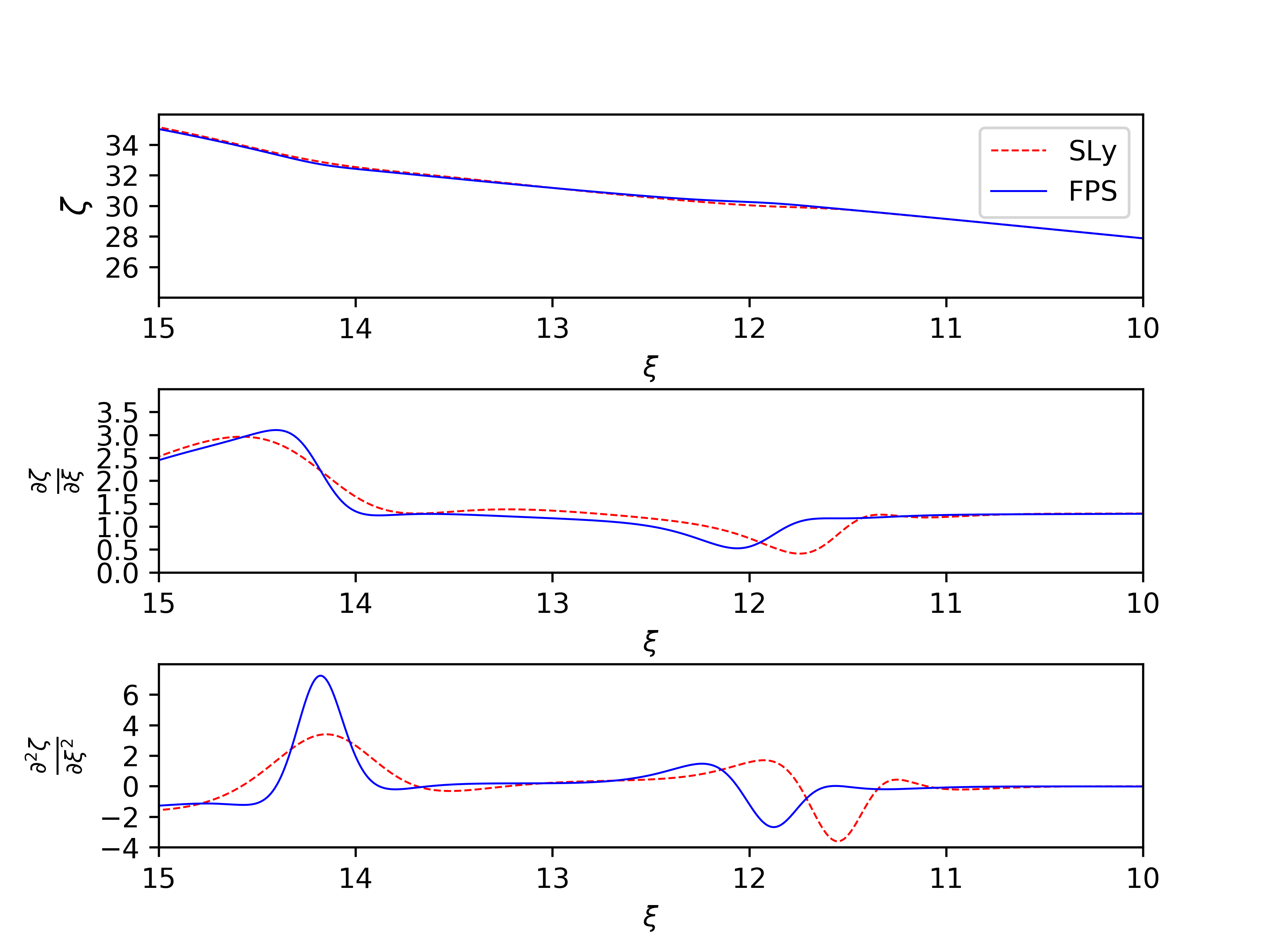}
  \caption{The Sly and FPS equation of state and their first and second derivatives.}
  \label{fig_EOS_derivatives}
\end{figure}

Additionally to the two analytic EoS above, we also studied neutron star solutions based on two other EoS in tabulated form. We choose EoS available in the CompOSE online service \cite{composeweb}, namely the so called HS(DD2) \cite{Hempel:2009mc,Typel:2009sy} and HS(TMA) \cite{Toki:1995ya,Hempel:2009mc} EoS. 
%and HS(TM1) \cite{Sugahara:1993wz,Hempel:2009mc} EoS. 
These two EoS are both based on the the statistical model presented in \cite{Hempel:2009mc}, which includes the contribution of nuclei, nucleons, electrons, positrons and photons (excluding neutrino contributions) and which requires as input the masses and binding energies of nuclei and an effective model for the nucleon-nucleon interaction, which is then treated in the relativistic mean-field (RMF) approximation. The two EoS thus differ in the different model used for that dynamical input. The HS(DD2) equation of state uses the density-dependent nuclear model called DD2 \cite{Typel:2009sy}, with the nuclei properties given by the FRDM model \cite{Moller:1996uf}. 
%The HS(TM1) equation of state uses the TM1 model %\cite{Sugahara:1993wz} for the nucleon interactions whilst the %masses of nuclei those in \cite{Geng:2005yu}. 
The HS(TMA) equation of state uses instead the TMA model \cite{Toki:1995ya}, with the masses of nuclei taken from \cite{Geng:2005yu}. 

The drawback of using tabulated data to solve the equations of stellar structure is that interpolation functions and numerical derivatives thereof must be used. This procedure is ambiguous (as one can use many different forms for the interpolating functions) and, as mentioned above, certain thermodynamic relations must be preserved by the derivatives of the EoS (see e.g. \cite{Haensel:2004nu}). This is a particularly difficult problem to tackle in the case of modified Palatini theories, since higher derivatives of the EoS are needed as compared to GR. 
% 
% The EoS tables can be generated by the \textsc{compose} program, which is provided at the CompOSE website [CITE WEBSITE?]. The pressure $P$ is directly calculated by the \textsc{compose} program, while for the energy density $\epsilon$ one needs to multiply the internal energy per baryon $\frac{E}{b}$ with the baryon number density $n_b$, i.e. $\epsilon = \frac{E}{b}n_b$. We calculated all quantities with \textsc{compose} in beta equilibrium at a temperature of 100 keV, with a log spaced grid with 10$^5$ grid points [EXPLAIN IT IN MORE DETAIL? MAYBE IN APPENDIX? OR IN A FOOTNOTE? OR IS IT OKAY THIS WAY?]. The derivatives of $\epsilon$ with respect to the pressure, which turn up in the stellar structure equations, have been calculated numerically with the \textit{numpy.gradient()} method, by applying it once for the first derivative and twice for the second. When calculating the derivatives we were only taking every 500$^{th}$ grid point in order to smooth out discontinuities in the EoS tables.
Indeed, in the stellar structure equations for Palatini $f(R)$ and $f(R,Q)$ the first and second derivative of the density $\rho$ with respect to the pressure $P$ appear. For the analytically given EoS above this is relatively straightforward. Remembering that $P(\rho)=10^{\zeta(\xi(\rho))}=10^{\zeta(\text{log}_{10}(\rho))}$ we have
\begin{equation}\label{eq_dp}
\frac{dP(\rho)}{d\rho}=\frac{dP}{d\zeta}\frac{d\zeta}{d\xi}\frac{d\xi}{d\rho} = \frac{P}{\rho}\frac{d\zeta}{d\xi}~,
\end{equation}
and
\begin{equation}\label{eq_d2p}
\frac{d^2P}{d\rho^2}=\frac{d}{d\rho}\left(\frac{dP}{d\rho}\right)=\frac{dP}{d\rho}\frac{1}{\rho}\frac{d\zeta}{d\xi}-\frac{P}{\rho^2}\frac{d\zeta}{d\xi}+\frac{P}{\rho^2}\frac{d^2\zeta}{d\xi^2}~,
\end{equation}
in order to arrive at the first and second derivative of the density with respect to the pressure we now have to invert Eqs.~\ref{eq_dp} and \ref{eq_d2p}. One gets
\begin{equation}\label{eq_invert_dp}
\frac{d\rho}{dP}=\frac{1}{\frac{dP}{d\rho}}~,
\end{equation}
\begin{equation}\label{eq_invert_d2p}
\frac{d^2\rho}{dP^2}=-\frac{\frac{d^2P}{d\rho^2}}{\left(\frac{dP}{d\rho}\right)^3}~.
\end{equation}
The result for the parametrization of the SLy and FPS EoS together with the first and second derivative of the parametrisations are shown in Fig.~\ref{fig_EOS_derivatives}. 
% A similar procedure must be applied to the EoS given in %form of tabulated data, but the different ways in which %the data can be interpolated and hence derivatives can be %taken has an impact in the results, as we discuss below.
For tabulated EoS data one can either apply a similar procedure as above after defining appropriate interpolating functions or simply calculate the derivatives numerically, as we did. When calculating the derivatives numerically, the different ways in which the data can be interpolated and hence derivatives can be taken has an impact on the results, as we discuss below.

\section{Results}\label{sec_results}

The mass and radius of a star are obtained by integrating from inside out the stellar structure equations after fixing a central density and until the pressure reaches a pre-determined threshold. In our calculations, the threshold was set to  $P=P_c \cdot 10^{-12}$ where $P_c$ is the initial central pressure (determined by the EoS from the central density) in case of the analytic EoS. For tabulated EoS we stopped the integration when reaching the last entry of the EoS tables. We considered central densities in  the range $\xi = [14.4,16]$ (at $r=0.01$mm for numerical stability). The integration was performed using a fourth-order Runge-Kutta method with a fixed stepsize of $1$m, since after exploring several adaptive-step methods we concluded that a fixed step was accurate enough and sped up the calculations.

A problem arising whenever a modified theory of gravity is used, is to determine the relative strength of the additional terms of the Lagrangian. Since in the present paper we use theories of the type $f(R)=R+\alpha R^2$ and $f(R,Q)=R+\alpha R^2 + \beta Q$, the problem reduces to finding reasonable values for $\alpha$ and $\beta$, consistent with phenomenology. However until now no solid experimental bounds on the values of $\alpha$ and $\beta$ exist for theories in the Palatini formalism (see \cite{Sotiriou:2005cd} for a study in Palatini). There exist, however, more restrictive experimental bounds on $\alpha$ for theories in the metric formalism (where only the metric is considered a dynamical degree of freedom). The Gravity Probe B experiment constrains $\alpha$ to $\alpha \lesssim 5 \cdot 10^{15}$ cm$^2$, while the constraint coming the Pulsar B in PSR J0737-3039 is four orders of magnitude higher \cite{Naf:2010zy}. In \cite{Naf:2010zy} they derived an even more stringent constraint on $\alpha$ from the E\"{o}t-Wash experiment, constraining $\alpha$ to $\alpha \lesssim 10^{-6}$ cm$^2$. However, this bound was derived in the low curvature regime of a laboratory on earth. Since possible modifications to GR will likely only be relevant in the high curvature regime, we do not take the bound from this experiment into consideration. Recently another bound for $\alpha$ has been derived by calculating the stability of stars in metric $f(R)=R+\alpha R^2$ gravity \cite{Pretel:2020rqx}. Using polytropic EoS and looking for the maximum value of $\alpha$ which still fulfills the stability criteria they found $\alpha \lesssim 2.4 \cdot 10^8$ cm$^2$. 
%What still remains to be done is to use realistic neutron star EoS, like the analytic SLy and FPS EoS, in order to derive the bound on $\alpha$, since this might change the value. 
To the best of our knowledge, no bounds on $\beta$ exist at the moment.

Since the dynamics of metric and Palatini theories can be very different, the bounds discussed above may not apply in our case. Lacking a thorough analysis along the lines of \cite{Naf:2010zy,Pretel:2020rqx} for Palatini theories, we used values for $\alpha$ and $\beta$ for which an appreciable change in the mass-radius curve was obtained and that, for the case of $\alpha$, are still within the bounds obtained in the metric formalism and in \cite{Sotiriou:2005cd}.

\subsection{$R+\alpha R^2$ case}\label{sec:fR_results}

\begin{figure}[htbp]
\centering
\includegraphics[width=\linewidth]{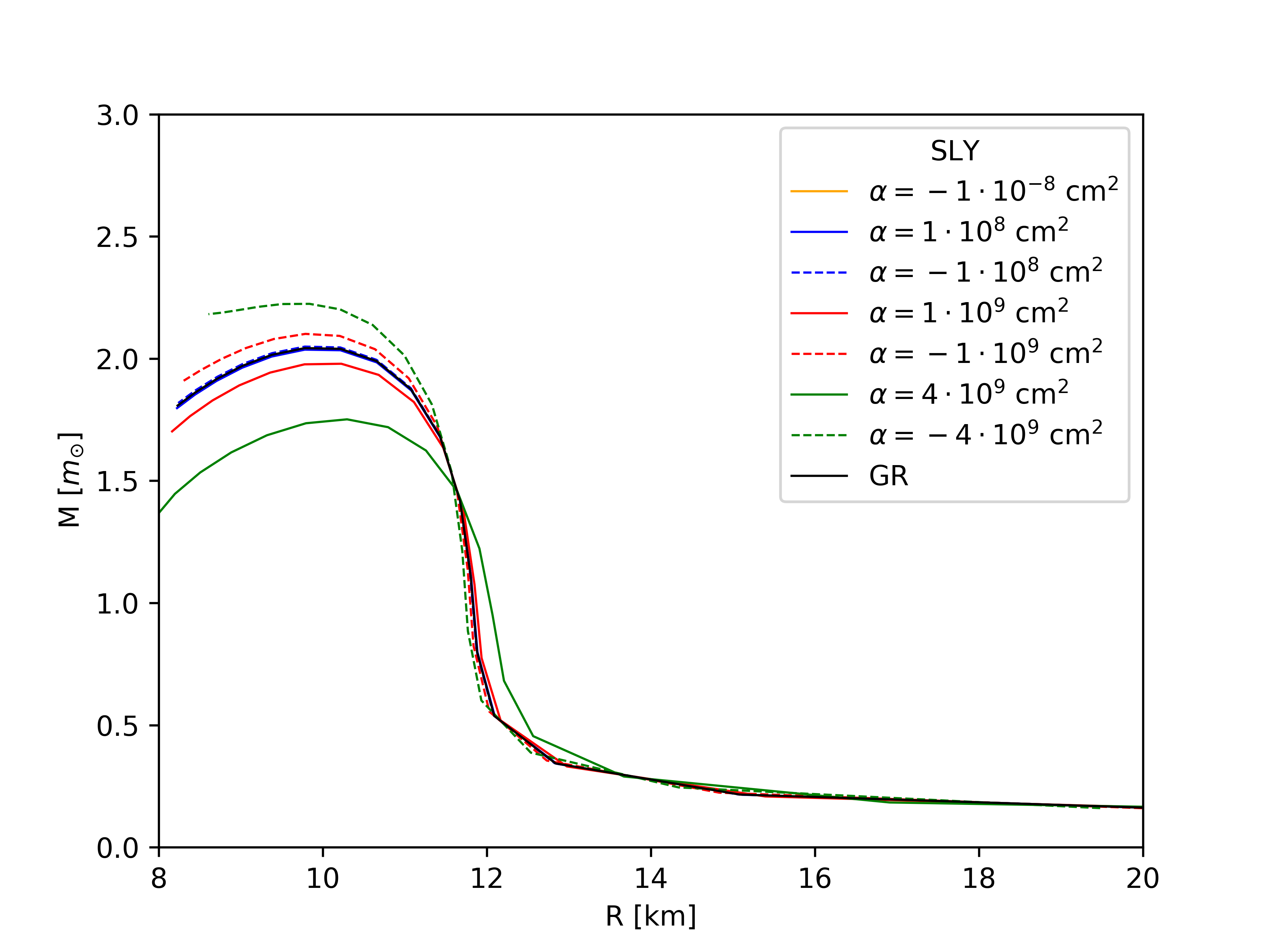}
\\
\centering
\includegraphics[width=\linewidth]{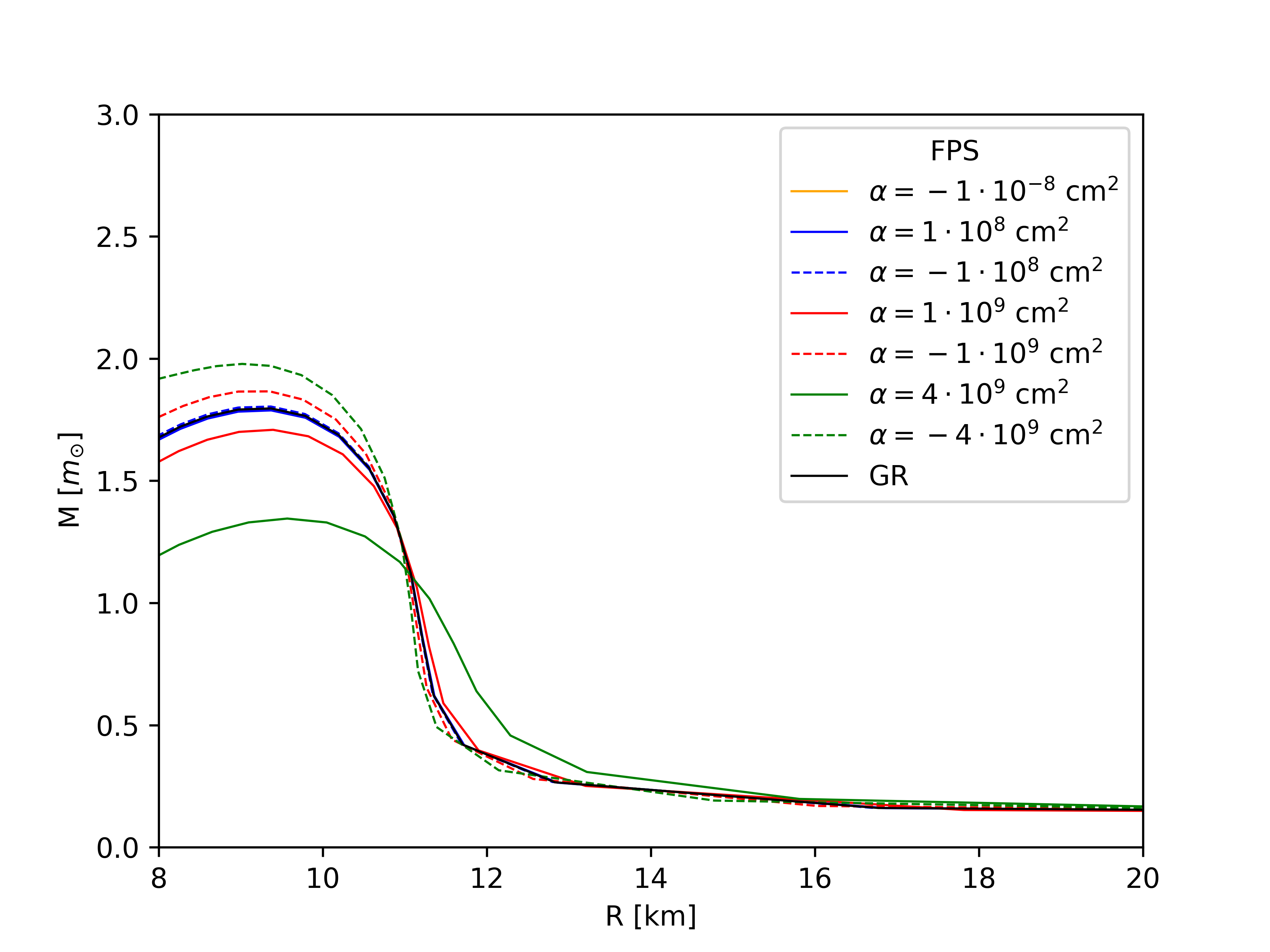}
\caption{Mass-radius relation for GR and $f(R)=R+\alpha R^2$, the SLy and FPS EoS and different values of the parameter $\alpha$.}
\label{fig_fR_sly_fps}
\end{figure}

We begin with the study of $f(R)=R+\alpha R^2$ theories and show the mass-radius relation for a range of positive and negative values of $\alpha$. Note that an analogous study was performed in \cite{Pannia:2016qbj} for the SLy and FPS EoS. However, our resuls are in contradiction with those presented in \cite{Pannia:2016qbj}. While in \cite{Pannia:2016qbj} the authors found only minor variations with respect to GR of the mass and radius of neutron stars, our calculations show instead a significant deviation from GR in some cases. We show in Fig.~\ref{fig_fR_sly_fps} our results for the mass-radius relation using the SLy and FPS EoS, compared to the corresponding GR result. As one can see, significant deviations from GR appear in the low-radius region even though smaller deviations are also observed for less compact stars. Interestingly, for negative values of $\alpha$, $f(R)=R+\alpha R^2$ theories generate heavier stars than GR and thus push the mass limit for the SLy and FPS EoS. 
%similarly huge variation of the mass and radius of the neutron %stars as one can find in \cite{Sylvester}, when using the analytic %SLy and FPS EoS. However, when using the tabulated EoS the results %depend on the particular EoS used. While some give results that %would be in agreement with \cite{Pannia:2016qbj}, i.e. they only give %minor variations compared to GR, one gives a variation that is %even bigger than the ones from the analytic EoS. As it turns out, %the magnitude of the variation not only depends on the parameter $%\alpha$ but also on the values of the first and second derivatives %of the EoS. This restricts the possible conclusions one can draw %from neutron star calculations about the viability of Palatini %$f(R)$ theories even further.

\begin{figure}[htbp]
\centering
\includegraphics[width=\linewidth]{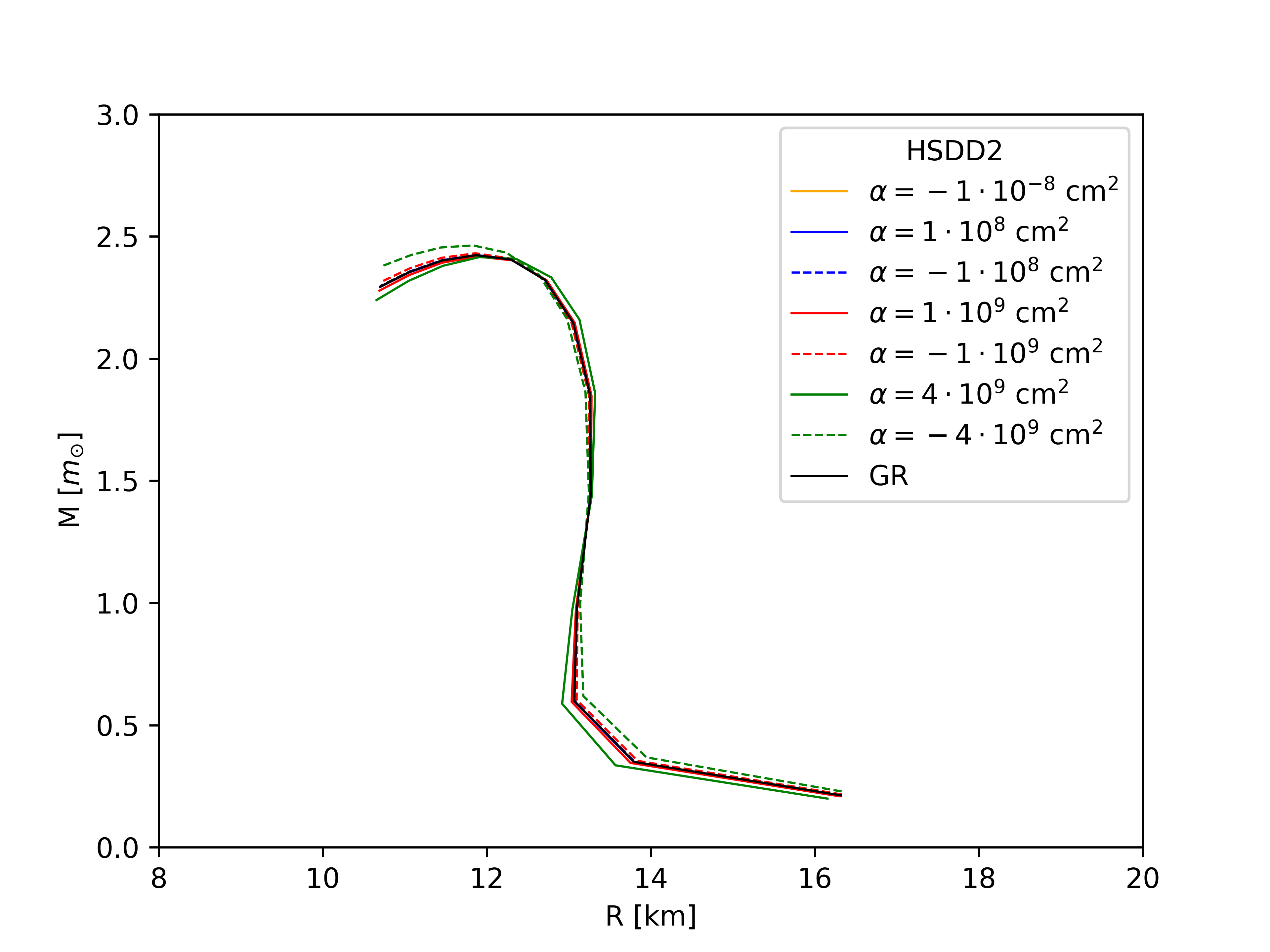}
\\
\centering
\includegraphics[width=\linewidth]{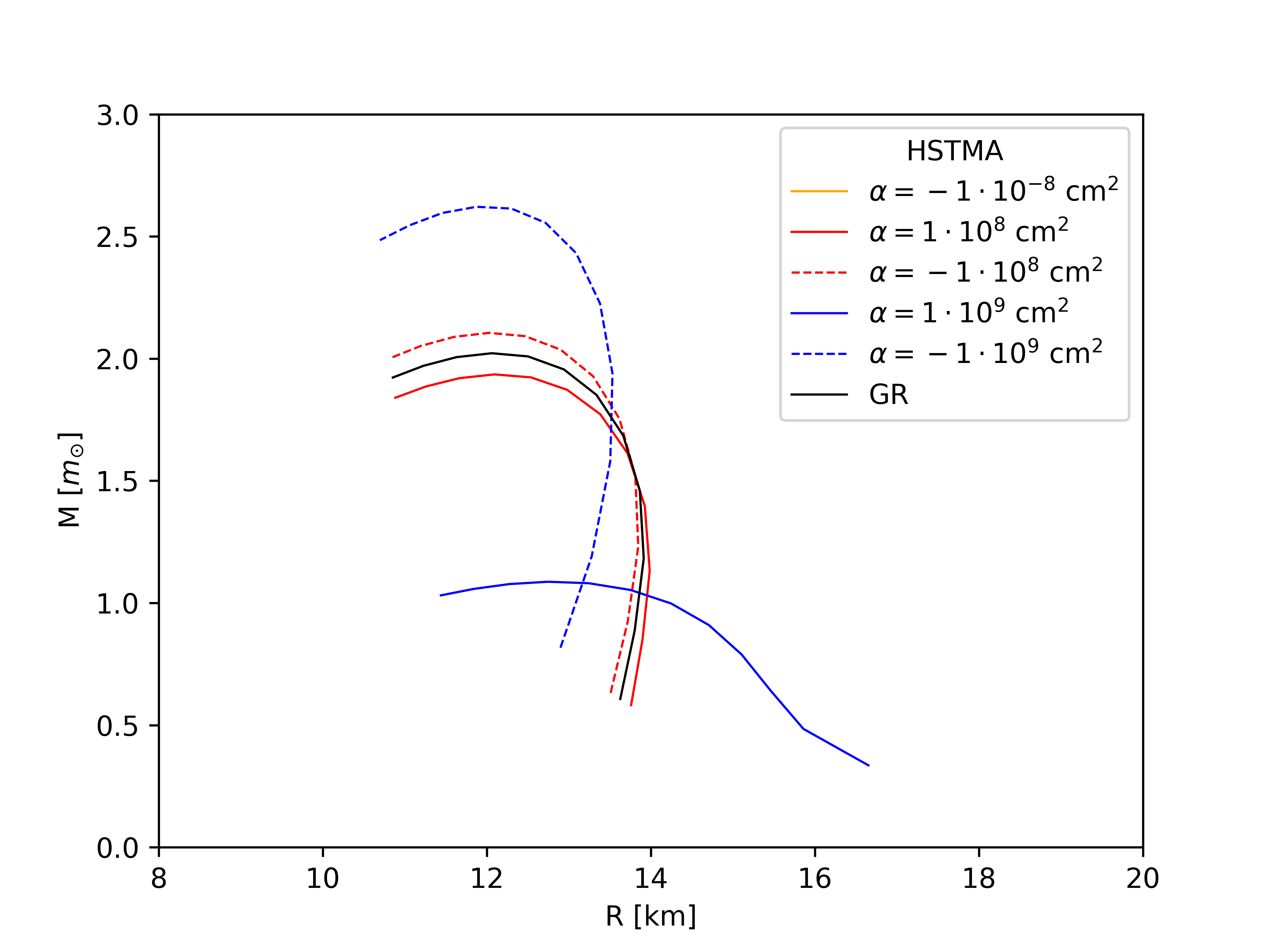}
\caption{Mass-radius relation for GR and $f(R)=R+\alpha R^2$, the HS(DD2) and HS(TMA) EoS and different values of the parameter $\alpha$.}
\label{fig_fR_hsdd2_hstma}
\end{figure}

\begin{figure}[htbp]
\centering
  \includegraphics[width=\linewidth]{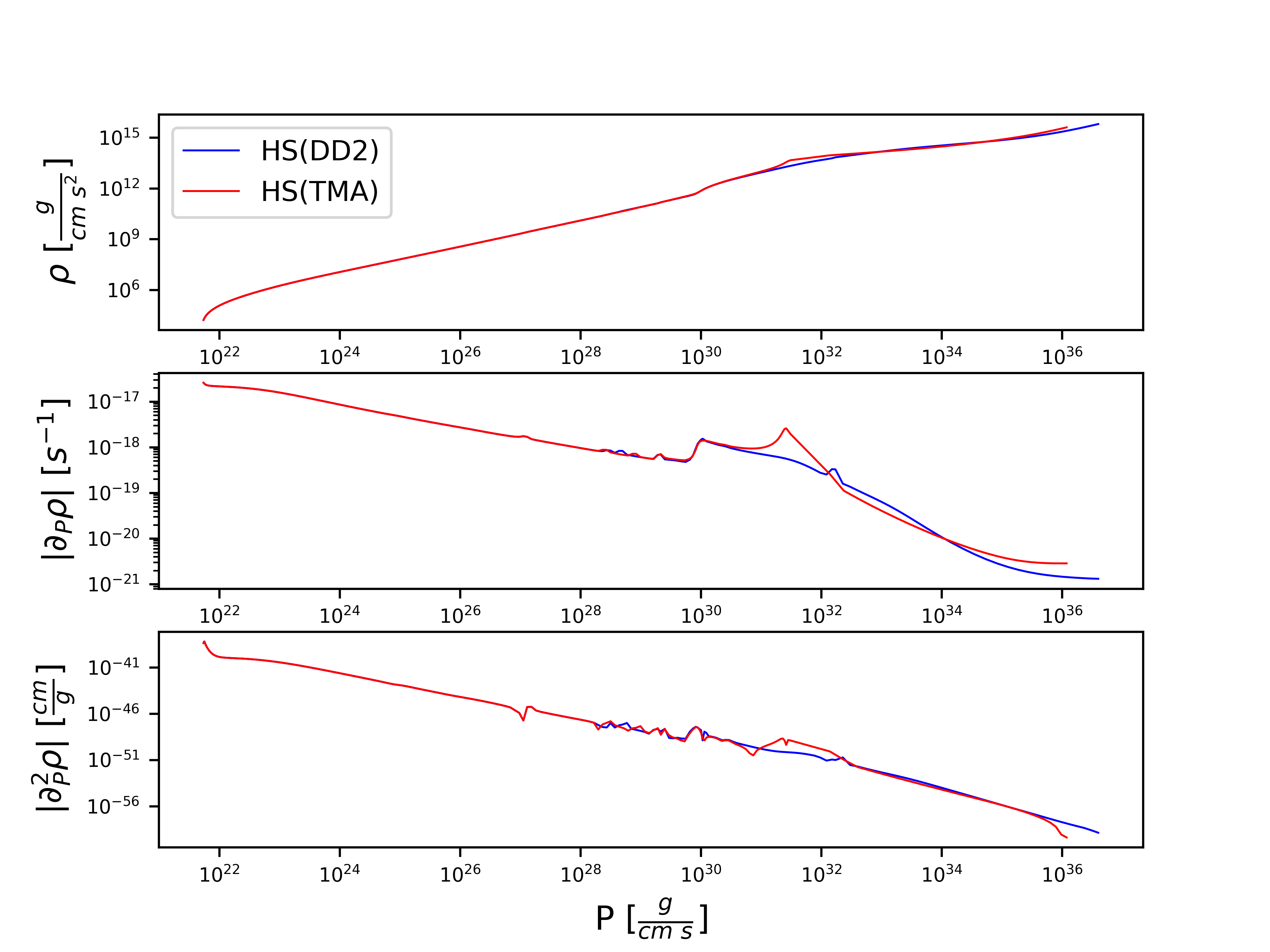}
  \caption{The EoS used in this paper (first panel) together with their first and second derivatives (second and third panel respectively).
}
  \label{fig_tabEOS_derivatives}
\end{figure}

The situation is less pronounced in the case of the tabulated HS(DD2) EoS, Fig.~\ref{fig_fR_hsdd2_hstma}. In this case, the differences in the mass-radius relation with respect to GR are still visible but are significantly smaller than in the previous case. What remains true is that negative values of $\alpha$ allow for a larger maximum mass. In the case of the HS(TMA) EoS, Fig.~\ref{fig_fR_hsdd2_hstma}, the deviations with respect to GR are dramatic for the largest values of $\alpha$ we used. This is certainly an unexpected behaviour that we were able to trace back to a small difference in the HS(TMA) EoS with respect to the other EoS we used\footnote{To confirm this hypothesis, we performed calculations using the HS(TM1) and SFHO EoS as well, which all show the same behaviour in their derivatives as HS(DD2) and do not show as large deviations from GR as HS(TMA).}, which induces a larger difference in the first and second derivative of the EoS. As can be seen in the first panel of Fig. \ref{fig_tabEOS_derivatives} at around $P=10^{32}\ \frac{g}{cm\ s}$ the HS(TMA) EoS differs slightly from all other EoS but the difference is amplified for the first and second derivatives  as can be seen in the second and third panel of Fig. \ref{fig_tabEOS_derivatives}.
Note finally, that the range of radii obtained in the case of tabulated EoS is limited by the maximum and minimum values of the data provided, unlike in the case where we used analytic expressions.

\begin{figure}[htbp]
  \centering
  \includegraphics[width=\linewidth]{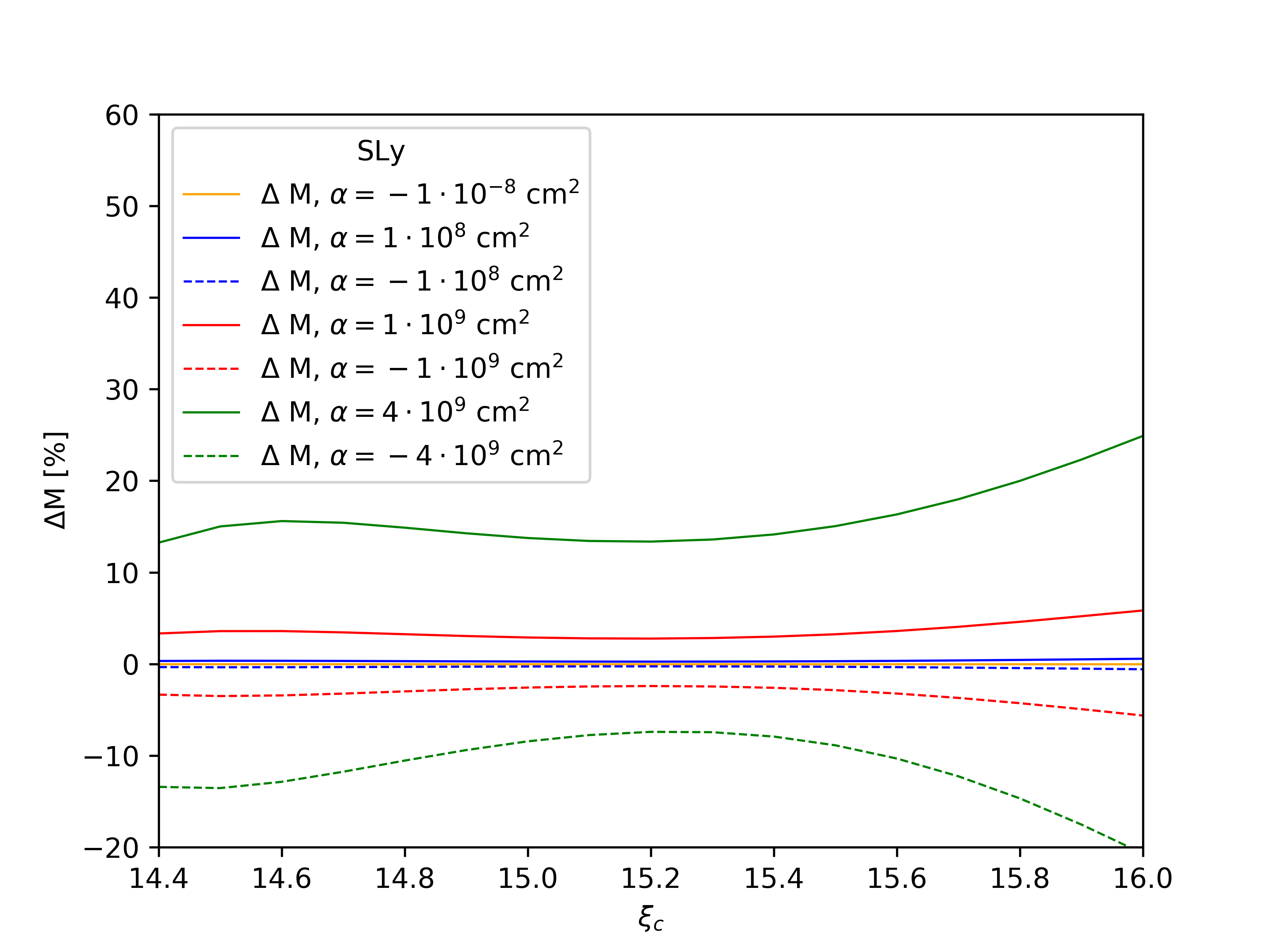}
\\
\centering
  \includegraphics[width=\linewidth]{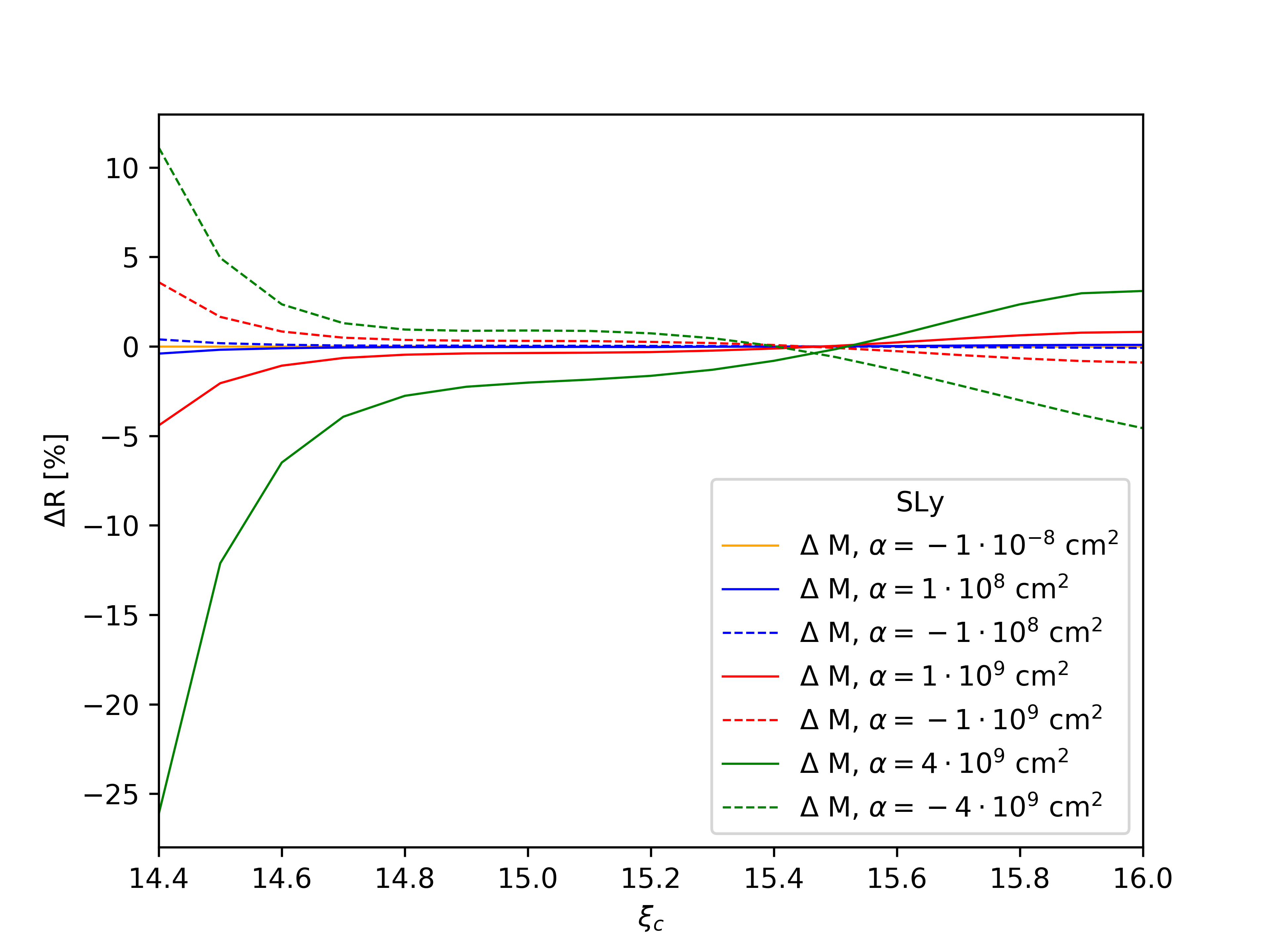}
\caption{Difference between the mass (upper panel) and the radius (lower panel) obtained, given a central density value $\xi$, with $f(R,Q)=R+\alpha R^2$ and GR, for the SLY EoS.}
\label{fig:DeltafR_SLY}
\end{figure}

\begin{figure}[htbp]
  \centering
  \includegraphics[width=\linewidth]{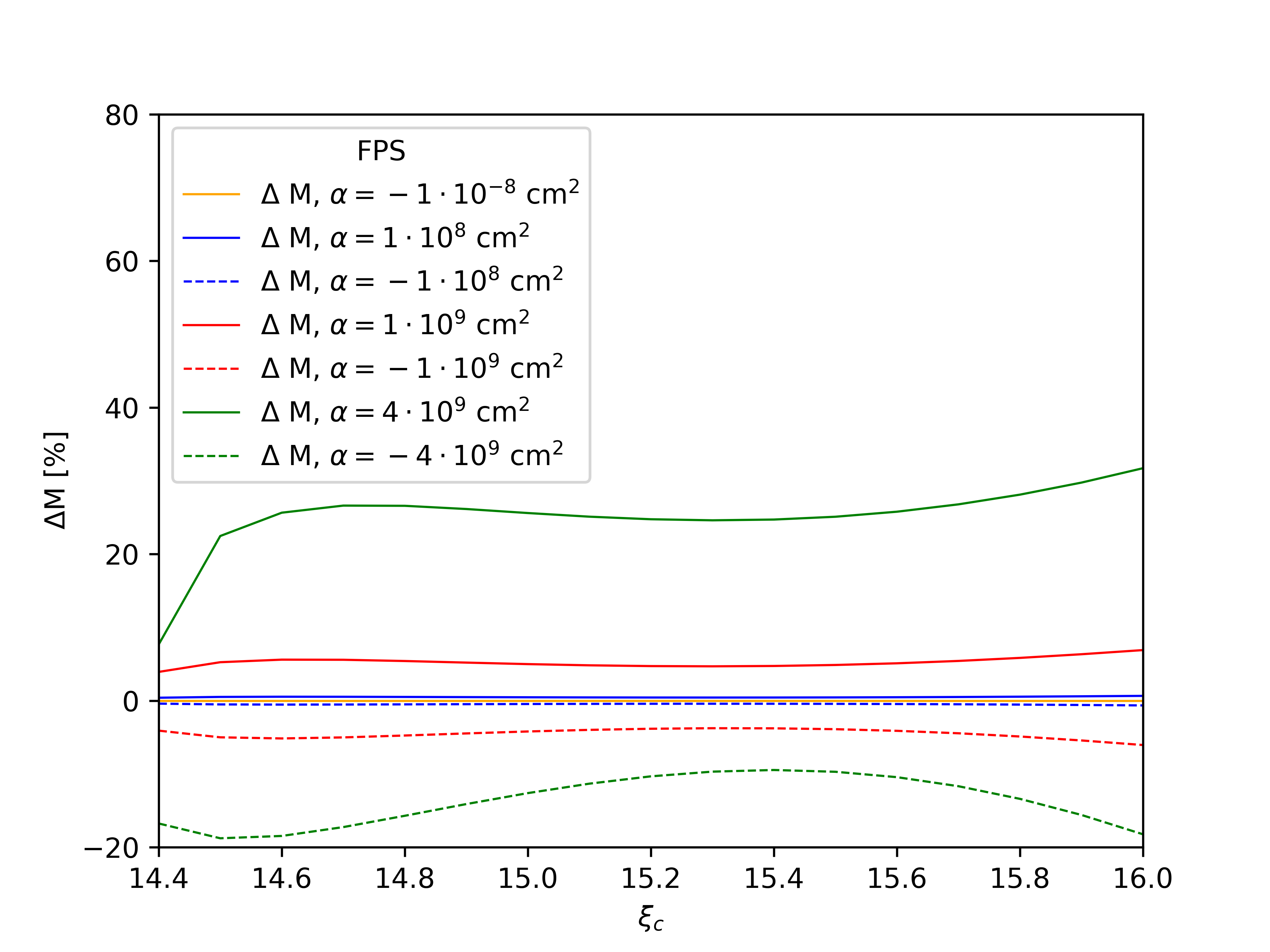}
\\
\centering
  \includegraphics[width=\linewidth]{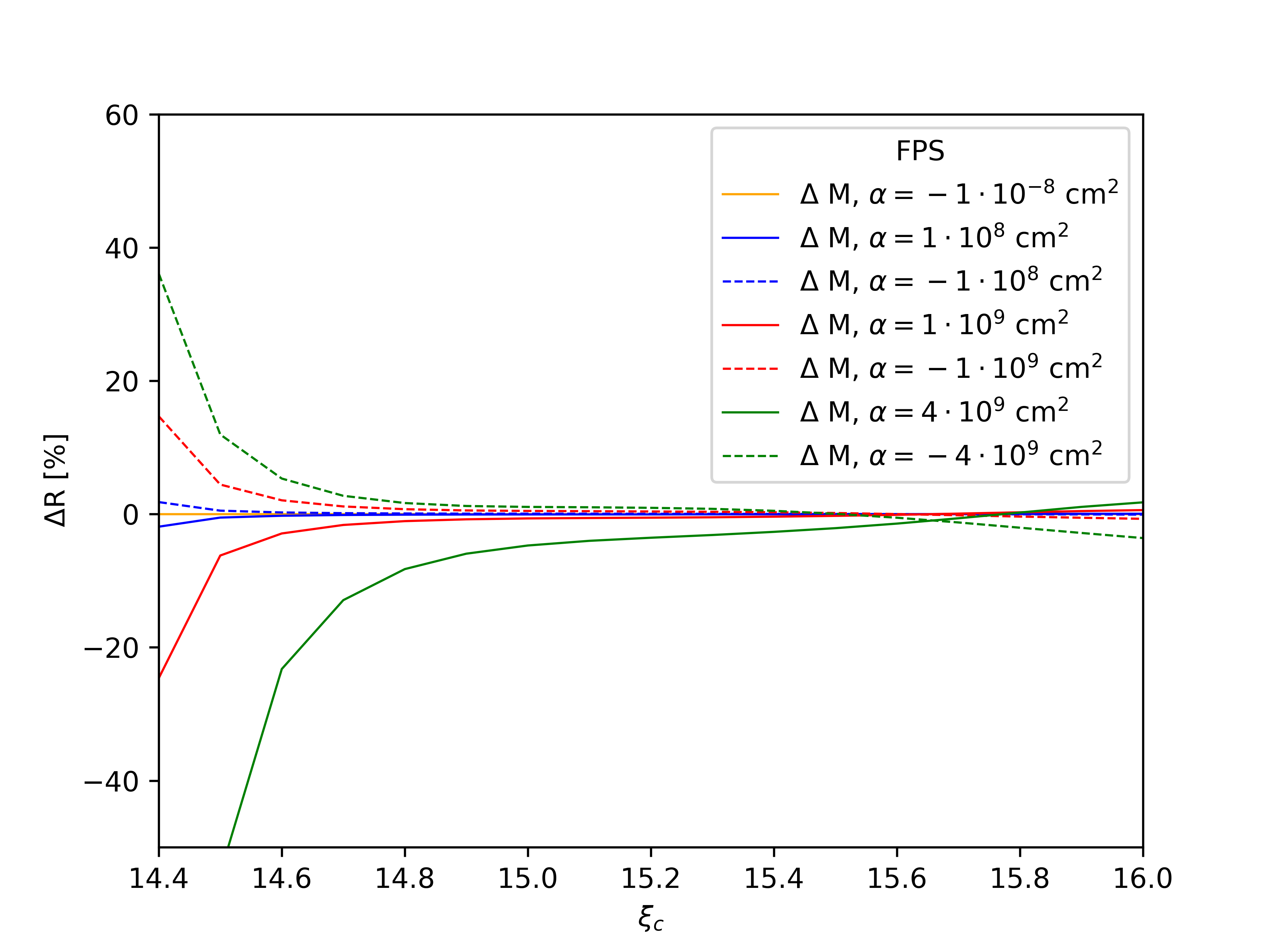}
\caption{Difference between the mass (upper panel) and the radius (lower panel) obtained, given a central density value $\xi$, with $f(R,Q)=R+\alpha R^2$ and GR, for the FPS EoS.}
\label{fig:DeltafR_FPS}
\end{figure}

\begin{figure}[htbp]
  \centering
  \includegraphics[width=\linewidth]{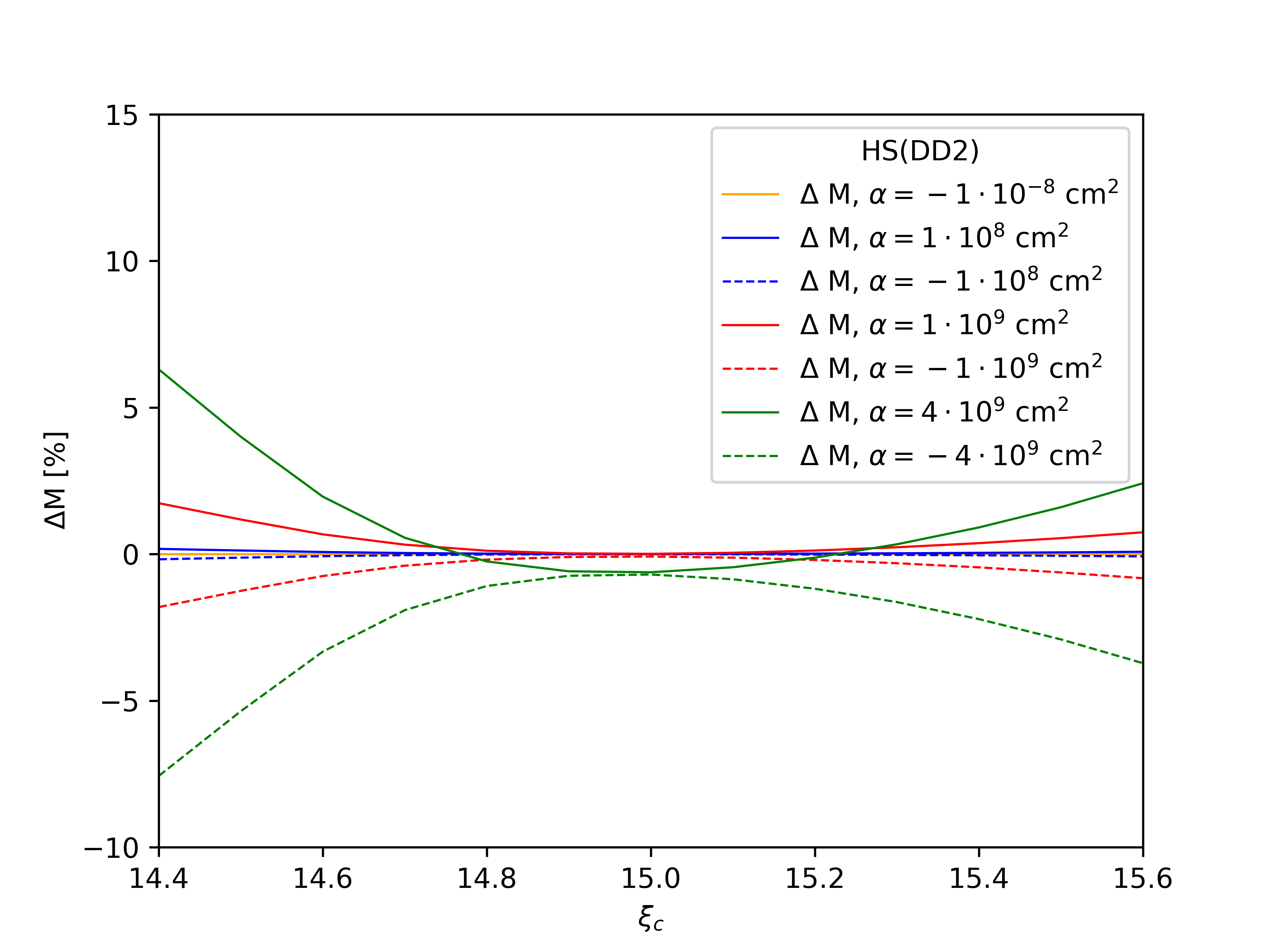}
\\
\centering
  \includegraphics[width=\linewidth]{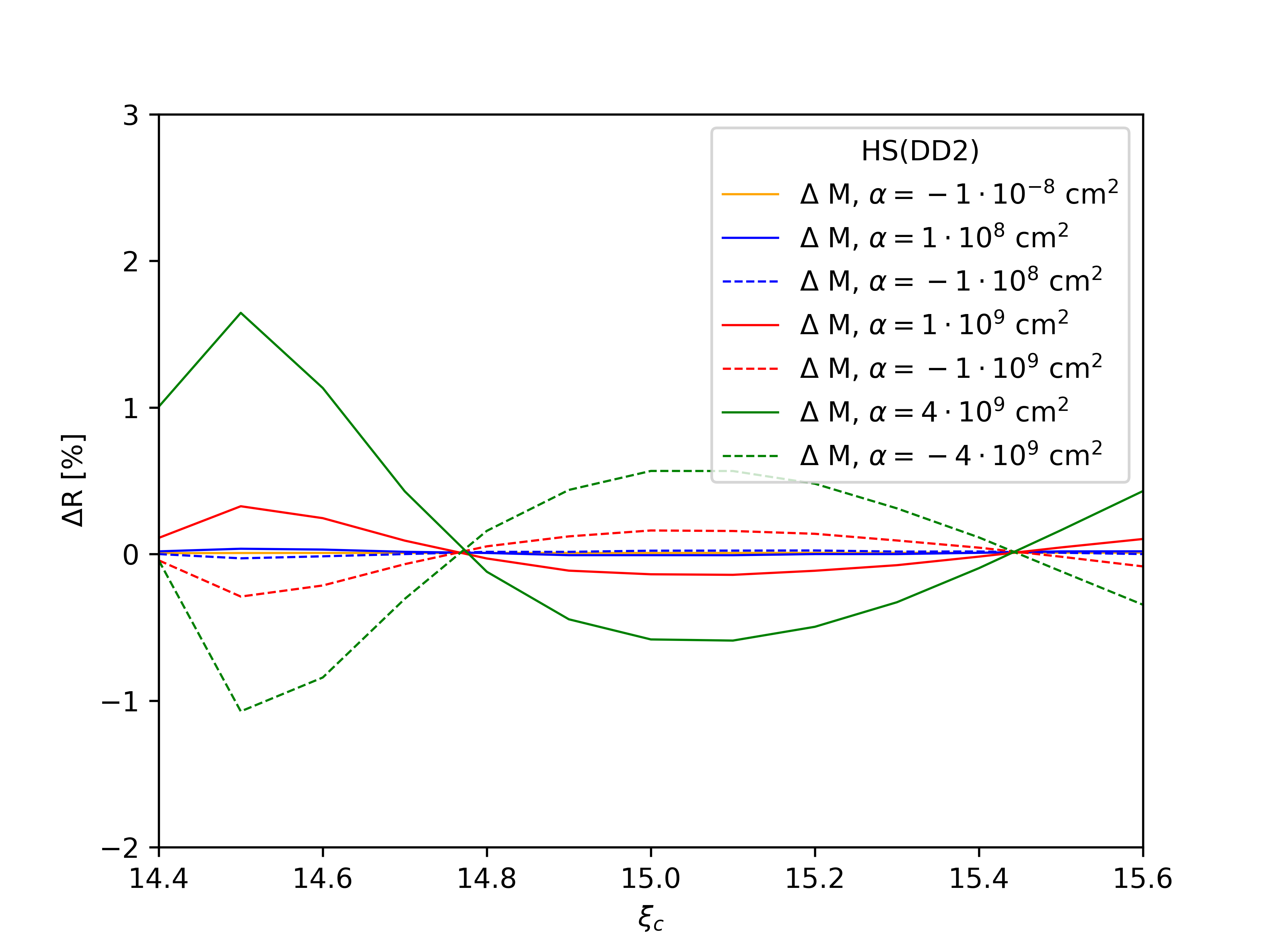}
\caption{Difference between the mass (upper panel) and the radius (lower panel) obtained, given a central density value $\xi$, with $f(R,Q)=R+\alpha R^2$ and GR, for the HS(DD2) EoS.}
\label{fig:DeltafR_HSDD2}
\end{figure}

\begin{figure}[htbp]
  \centering
  \includegraphics[width=\linewidth]{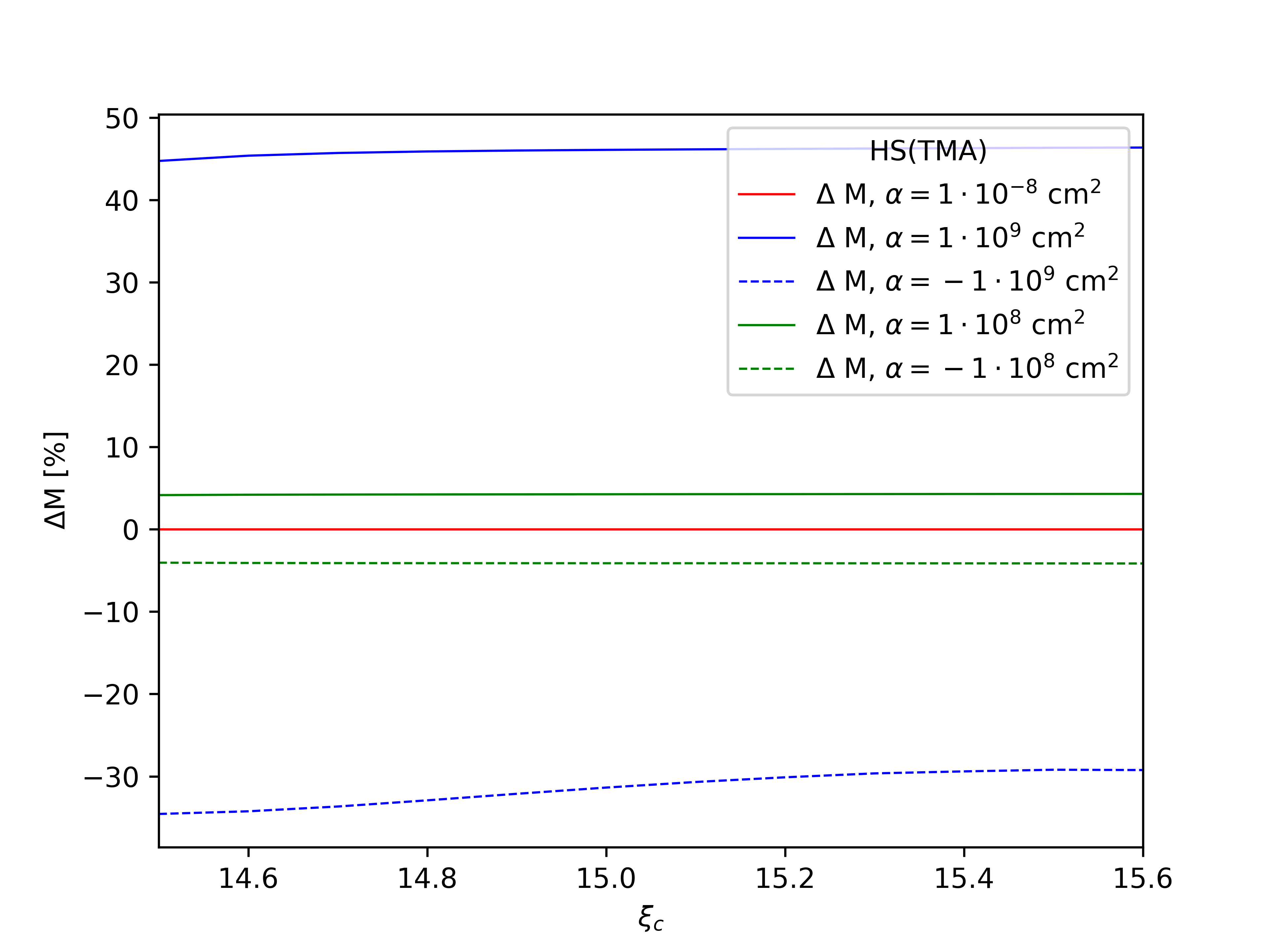}
\\
\centering
  \includegraphics[width=\linewidth]{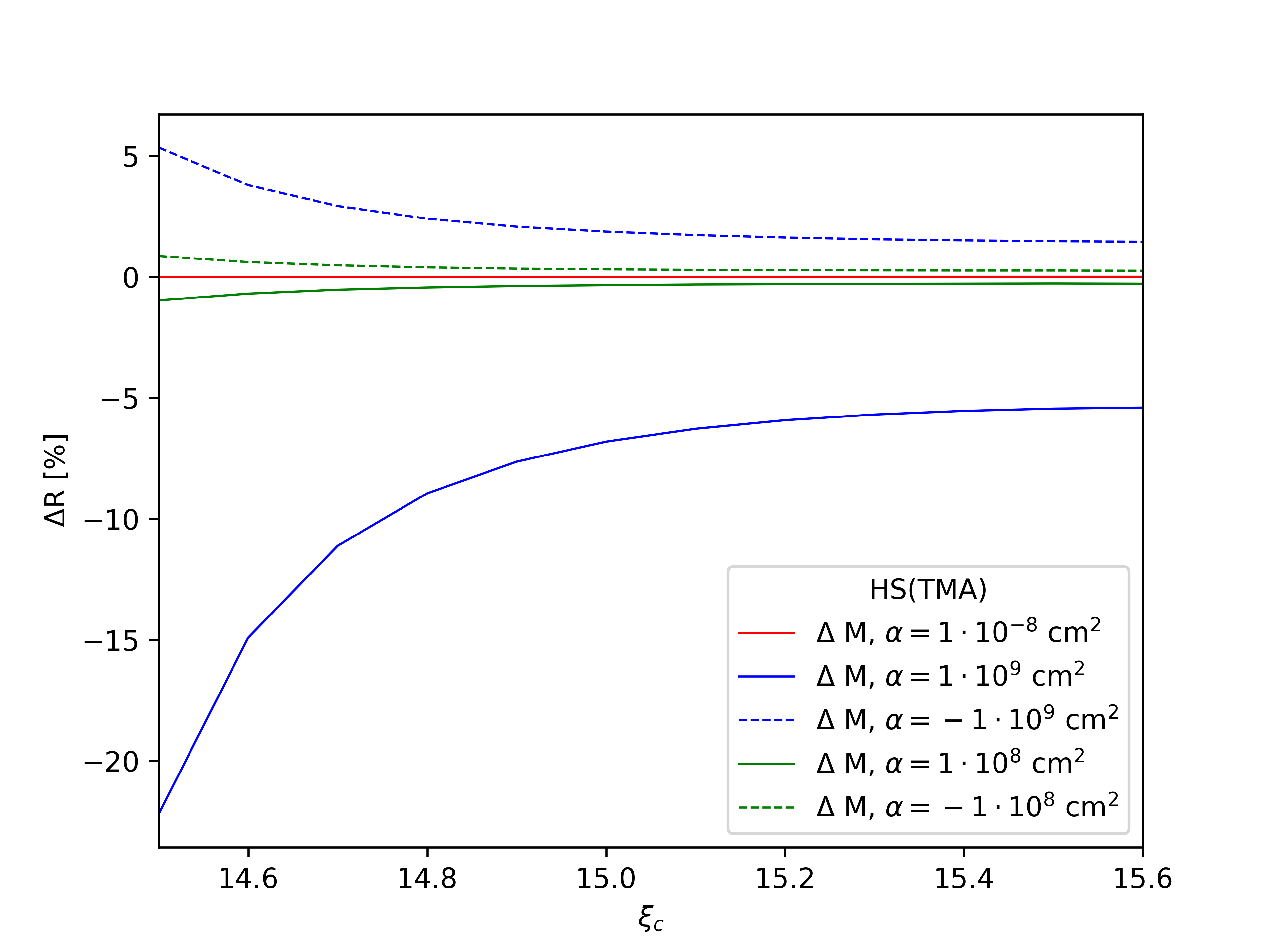}
\caption{Difference between the mass (upper panel) and the radius (lower panel) obtained, given a central density value $\xi$, with $f(R,Q)=R+\alpha R^2$ and GR, for the HS(TDA) EoS.}
\label{fig:DeltafR_HSTMA}
\end{figure}

In addition to the mass-radius plots, it is instructive to analyse the results as in Figs.~\ref{fig:DeltafR_SLY}--\ref{fig:DeltafR_HSTMA}, were we show the difference in percentage between the $f(R)$ and the GR solutions, for the mass and the radius of a star separately. In this way we observe that the effect of the $R^2$ term on the radius of stars is generally smaller than on the mass, except for the lowest central-density regions. The shift in the mass is thus the main responsible of the change of the mass-radius ratios discussed above. The changes, moreover, are strongly dependent on the EoS; SLy, FPS and HS(TMA) exhibit large deviations with respect to GR, but the latter shows a qualitatively different behaviour as a function of the central density. For HS(DD2), the deviations with respect to GR are significantly smaller than for the other EoS, for both mass and radius.

\begin{figure}[htbp]
\centering
\includegraphics[width=\linewidth]{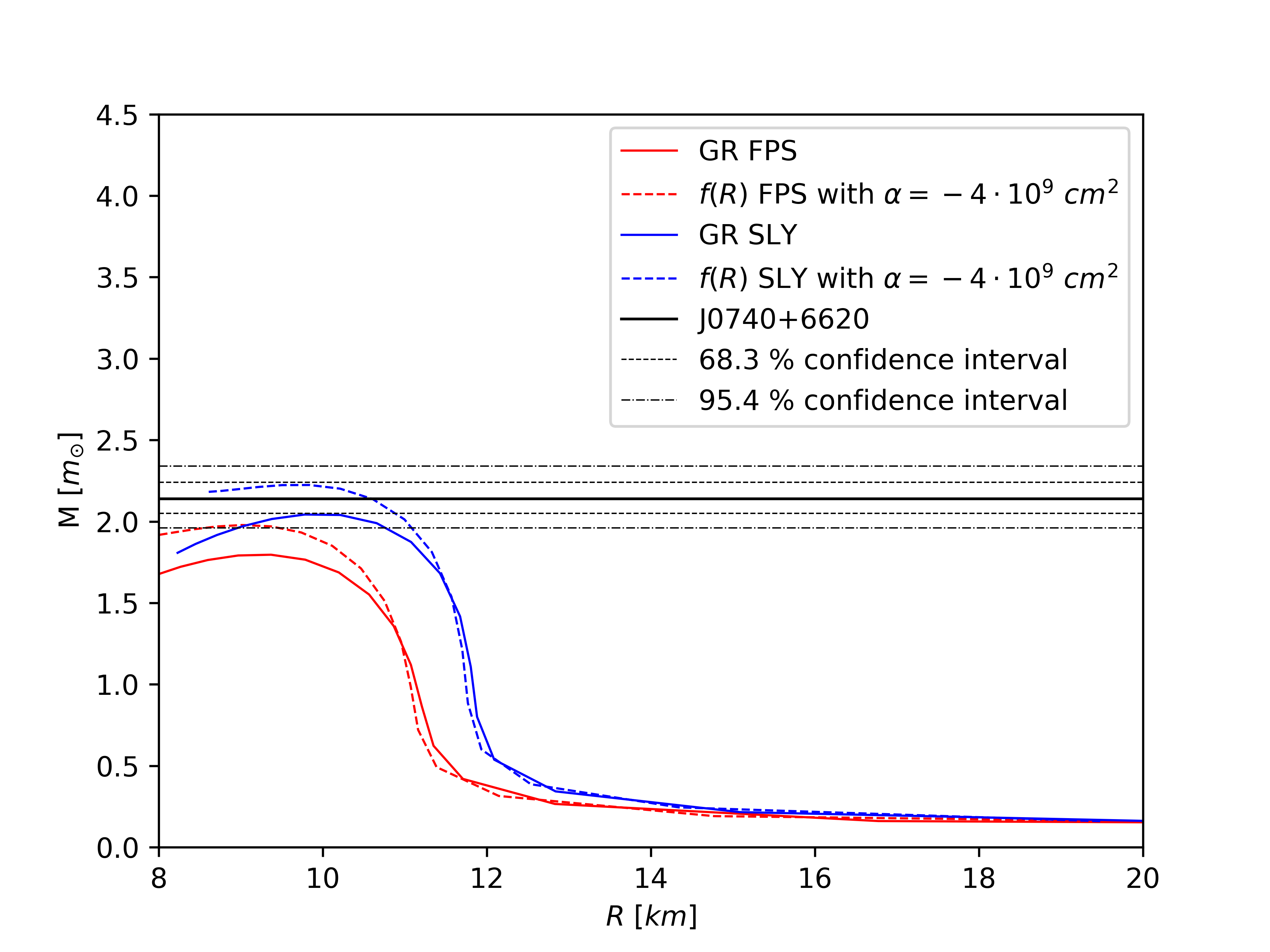}
\\
\centering
\includegraphics[width=\linewidth]{m_vs_r_compare_J0740_fR_v2.png}
\caption{Maximum mass allowed by an $f(R)=R+\alpha R^2$ theory with $\alpha=-4\cdot 10^{9}$cm$^2$ for the different EoS used in this paper, compared with the observed mass of the MSP J0740+6620 neutron star.}
\label{fig:StarJ0470_fR}
\end{figure}

It is clear from our discussion so far, that as long as the ambiguities in the EoS of neutron stars are not resolved, it is hardly possible to draw conclusions from mass and radius observations on the theory of gravity, as for a fixed $f(R)$ (here, for a given $\alpha$ value) the mass-radius relation changes dramatically for different EoS. However, it is still interesting to speculate how relevant would astrophysical observations be, if we knew the correct EoS of a neutron star. For example, the observation of the 2.2$M_\odot$ neutron star  MSP J0740+6620 \cite{Cromartie:2019kug} has been used as an argument to rule out certain EoS, including some used in this paper. This is can be seen in Fig.~\ref{fig:StarJ0470_fR}, where we show the constraints put by MSP J0740+6620 on EoS, assuming GR is the theory of gravity and also using $f(R)=R+\alpha R^2$ with $\alpha=-4\cdot 10^{9}$cm$^2$ (for which the increase of mass is the largest in our calculations). Clearly, some EoS can be \textit{rescued} adding an $R^2$ term to the theory of gravity. Reversing the argument, with increasing certainty in the EoS of neutron stars, observations could be used to constraint deviations of the theory of gravity from GR. 

%However, when looking at the results for tabulated EoS there is a %different picture. While one EoS shows a similar behavior as the %analytic EoS, i.e. it gives a huge variation compared to GR as %soon as the $f(R)$ modification is turned on, the other EoS are %more in line with the results of \cite{Pannia:2016qbj}. As can be seen %in Fig. \ref{fig_fR_hsdd2_hstma} the HS(TMA) EoS gives a huge %variation compared to GR already for values of $\alpha = 10^9$ %cm$^2$, while the HS(DD2) EoS only gives a small variation, which %is in line with the results of \cite{Pannia:2016qbj}. In Fig. %\ref{fig_fR_hstm1_sfho} one can see that also the HS(TM1) and SFHO EoS lead to only minor variations compared to GR.

\begin{figure}[htbp]
\centering
  \includegraphics[width=\linewidth]{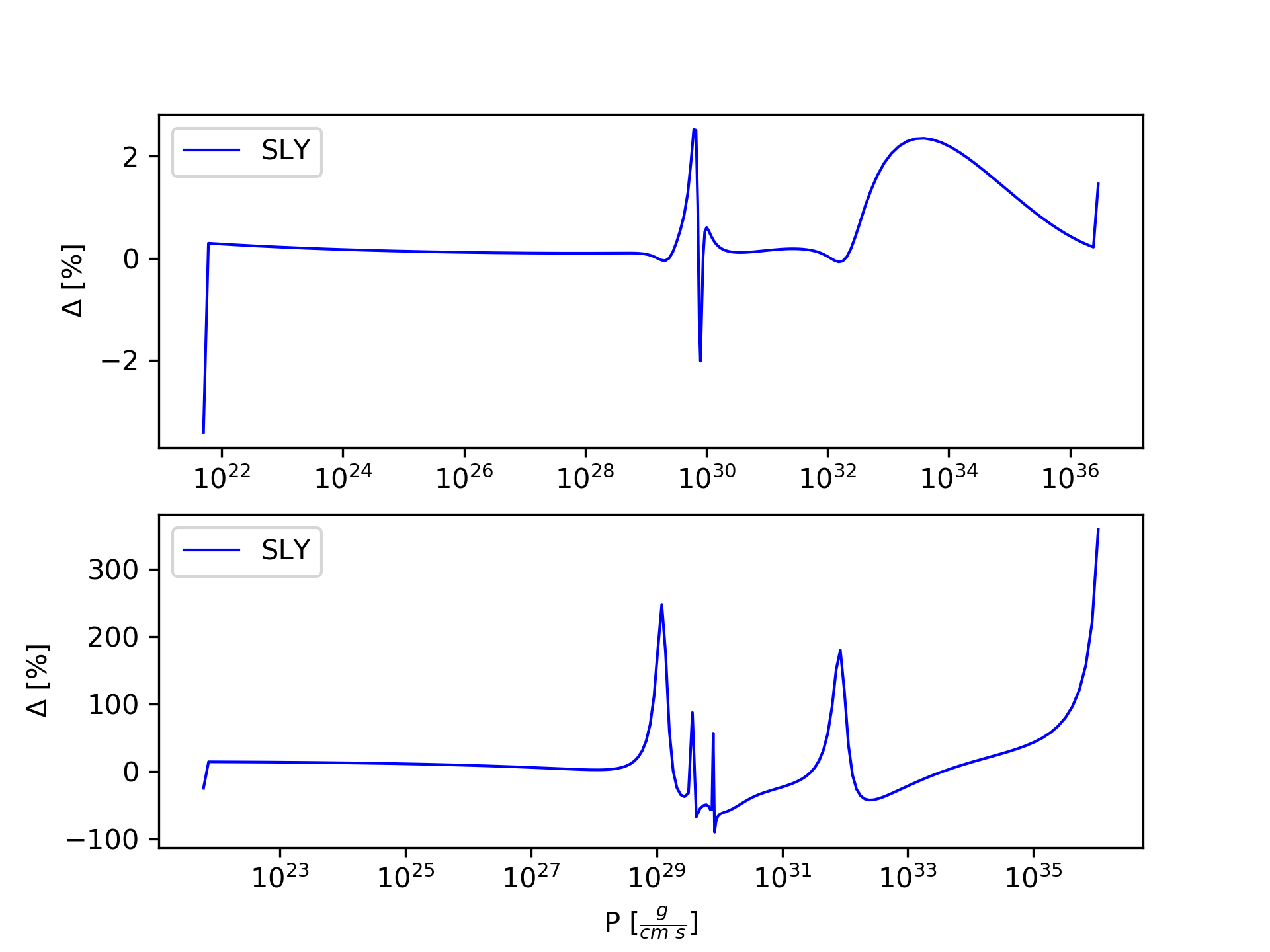}
  \caption{Difference between the exact and numeric derivative for the first (upper panel) and second derivative (lower panel) of the EoS.}
  \label{fig_EoS_dif}
\end{figure}

\begin{figure}[htbp]
  \centering
  \includegraphics[width=\linewidth]{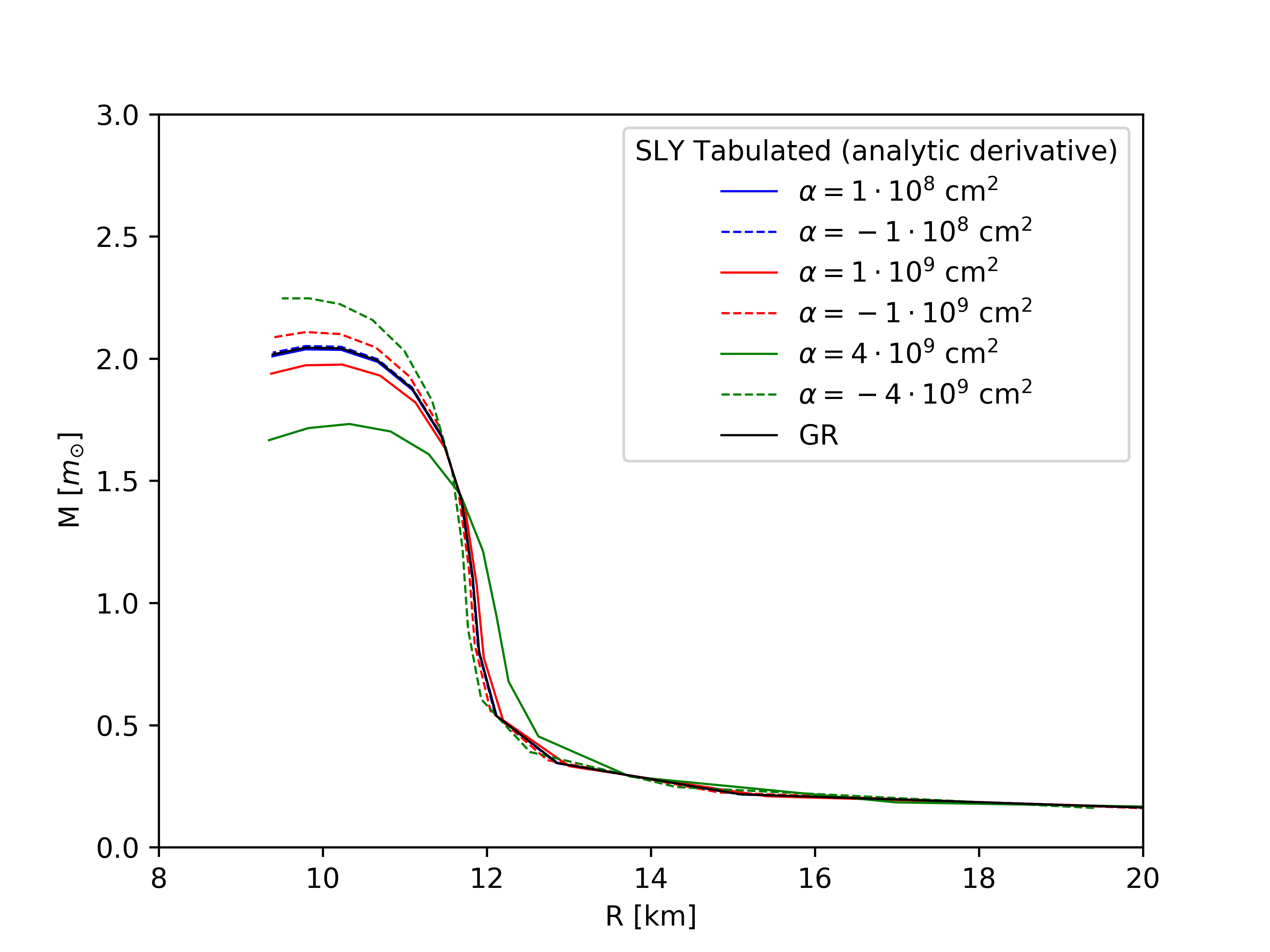}
\\
  \centering
  \includegraphics[width=\linewidth]{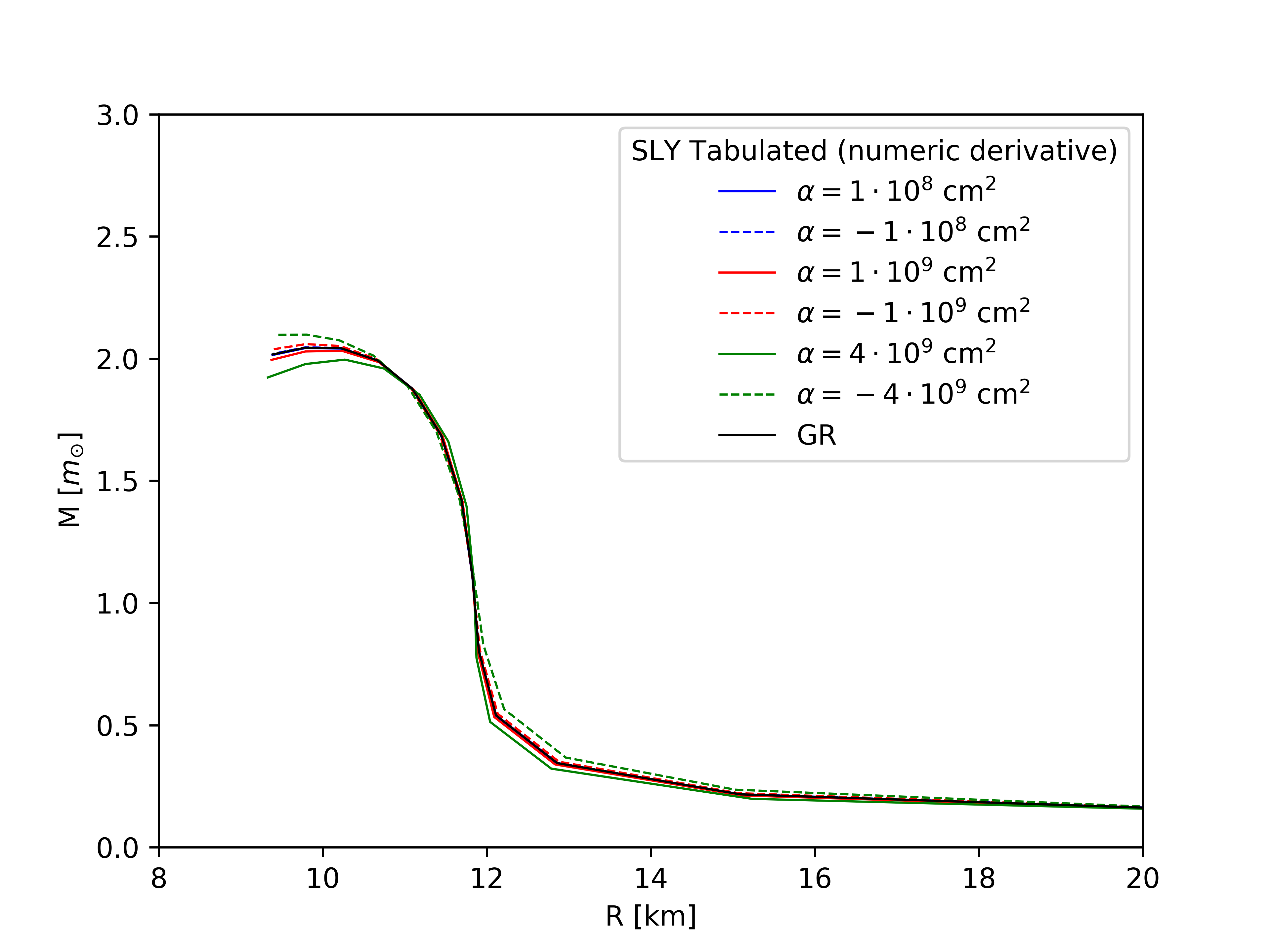}
\caption{Mass-radius relation for fake tabulated SLy EoS using the exact values for the derivative (left panel) and the values derived by the finite difference method (right panel).}
\label{fig_faketabulated_sly}
\end{figure}

%Before we proceed to present the results for $f(R,Q)$ theories
At this point, we would like to discuss in some detail the differences in the results induced by the different methods used to calculate derivatives of the EoS in the case of tabulated data. To this end, we repeated the calculations with the analytic EoS SLy, this time generating tabulated data from the analytic expressions as follows. In one case, the derivatives were calculated analytically as outlined in Sec.~\ref{sec_EOS} and separate data tables were generated from them; these values then had to be interpolated (in the same way as the EoS data is interpolated). As a second method we calculated the derivatives numerically with a finite difference approach, using second order central differences. In Fig. \ref{fig_EoS_dif} the difference between the derivatives using those methods can be seen. While the difference for the first derivative is rather small and never exceeds $4\%$, the second derivative shows very large differences, of about $200\%$ which rises up to $400\%$, at some points. To exemplify the impact of such differences in the mass-radius relations, we repeated the calculation for the SLy EoS with the different methods of calculating derivatives. The results are shown in Fig. \ref{fig_faketabulated_sly}. In the upper panel we show the calculation where the tabulated data for exact derivatives is interpolated, which clearly leads to the same results as when using the fully analytic version. However, in the lower panel, where the derivatives have been calculated with a finite difference method, shows only a minor difference between $f(R)$ and GR, in fact in agreement with the results published in \cite{Pannia:2016qbj}. This discussion points to a potential problem appearing when calculating with tabulated EoS and theories with high-order derivatives of matter; all  calculations that rely on numerical differentiation must  be taken with due caution.

\subsection{$R+\alpha R^2+\beta Q$ case}\label{sec:fRQ_results}

We proceed now to discuss the results for the $f(R,Q)=R+\alpha R^2+\beta Q$. As already indicated above, we consider values  of $\beta$ for which an appreciable change in the mass-radius curve is obtained. This means that, even the small changes in the stars masses that we observe, as discussed next, may be artificially large. 

\begin{figure}[htbp]
  \centering
  \includegraphics[width=\linewidth]{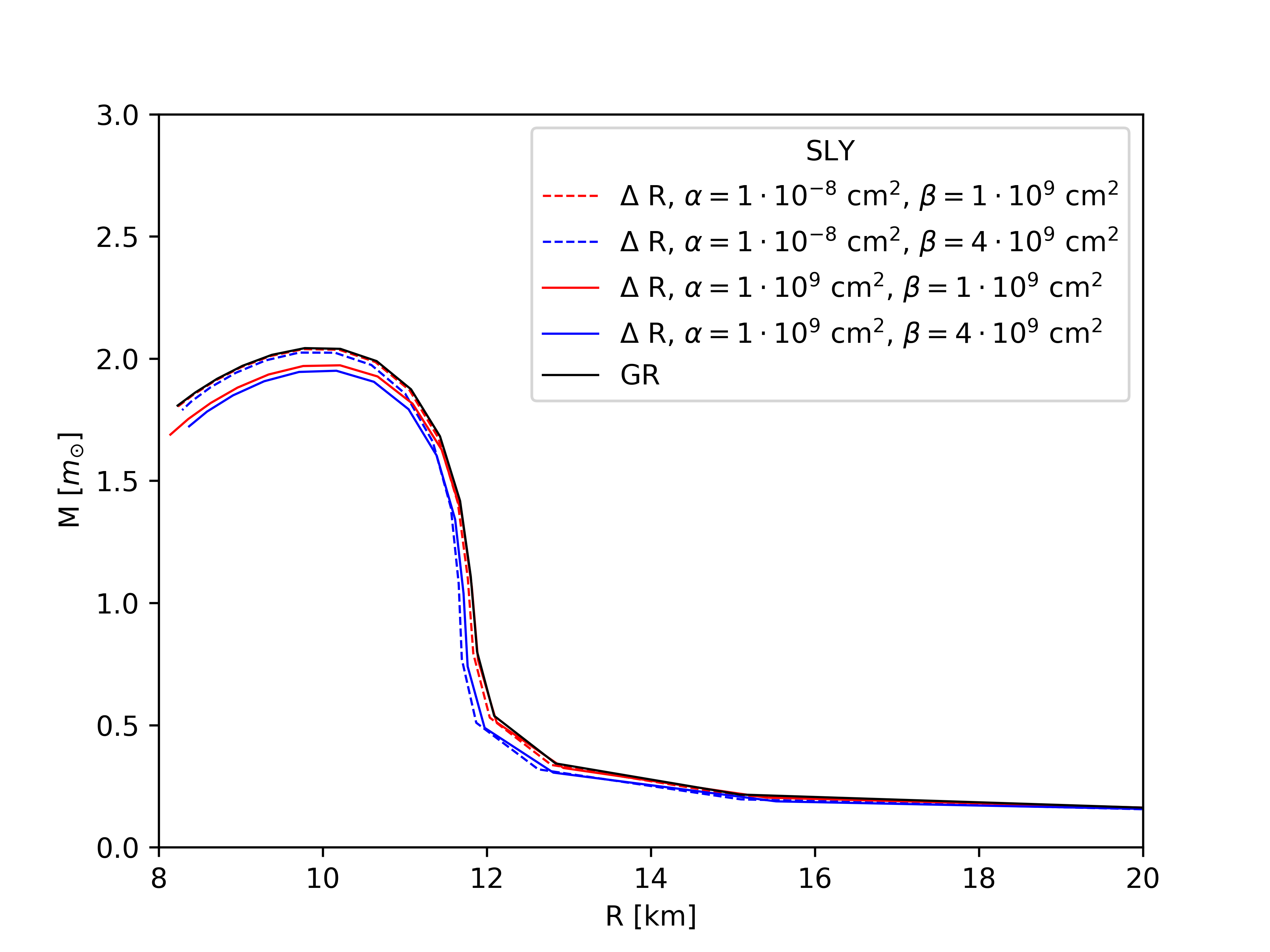}
\\
\centering
  \includegraphics[width=\linewidth]{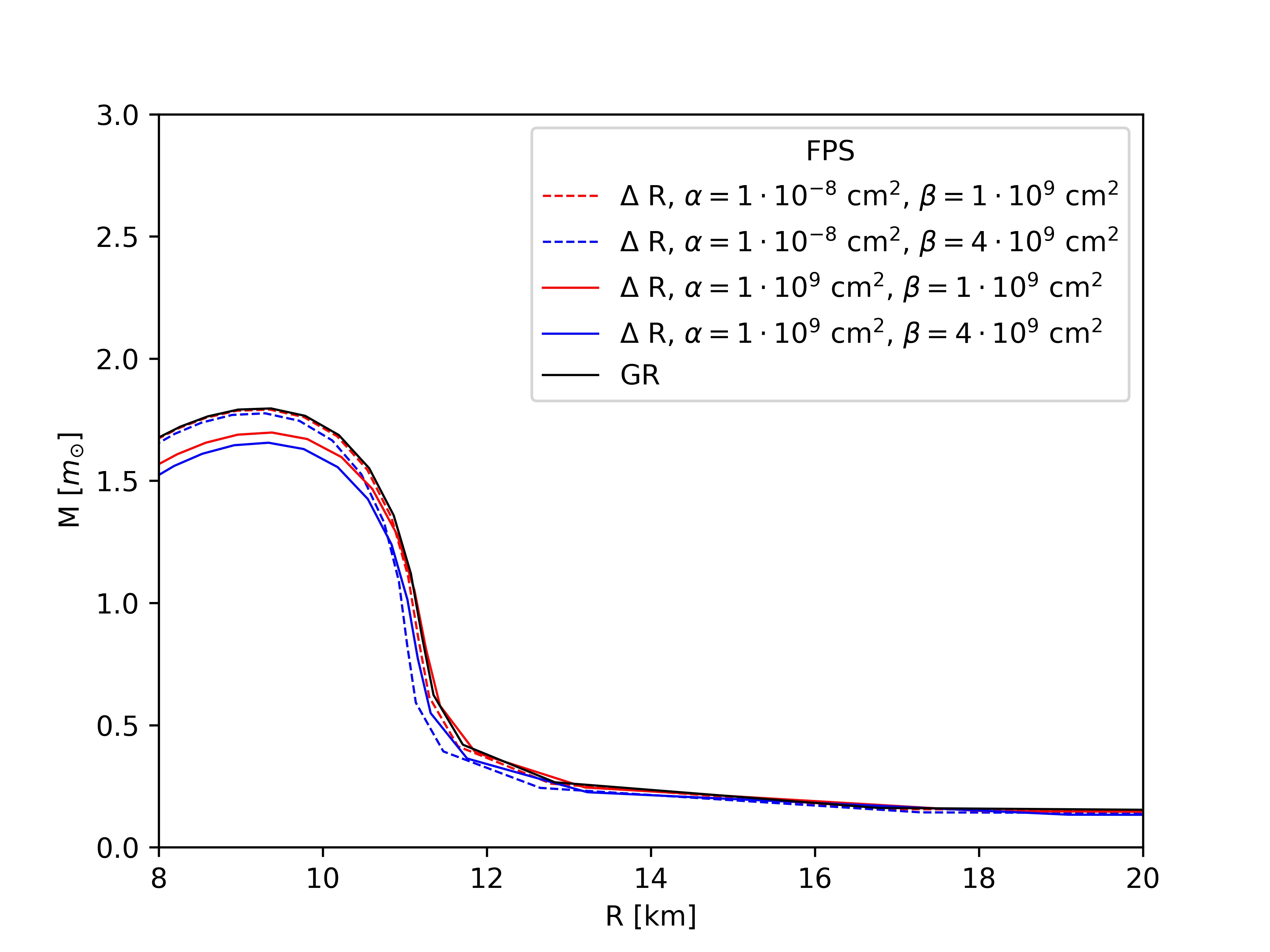}
\caption{Mass-radius relation for $f(R,Q)=R+\alpha R^2+\beta Q$ for SLY (upper panel) and FPS (lower panel) EoS.}
\label{fig:fRQ_analytic}
\end{figure}

\begin{figure}[htbp]
  \centering
  \includegraphics[width=\linewidth]{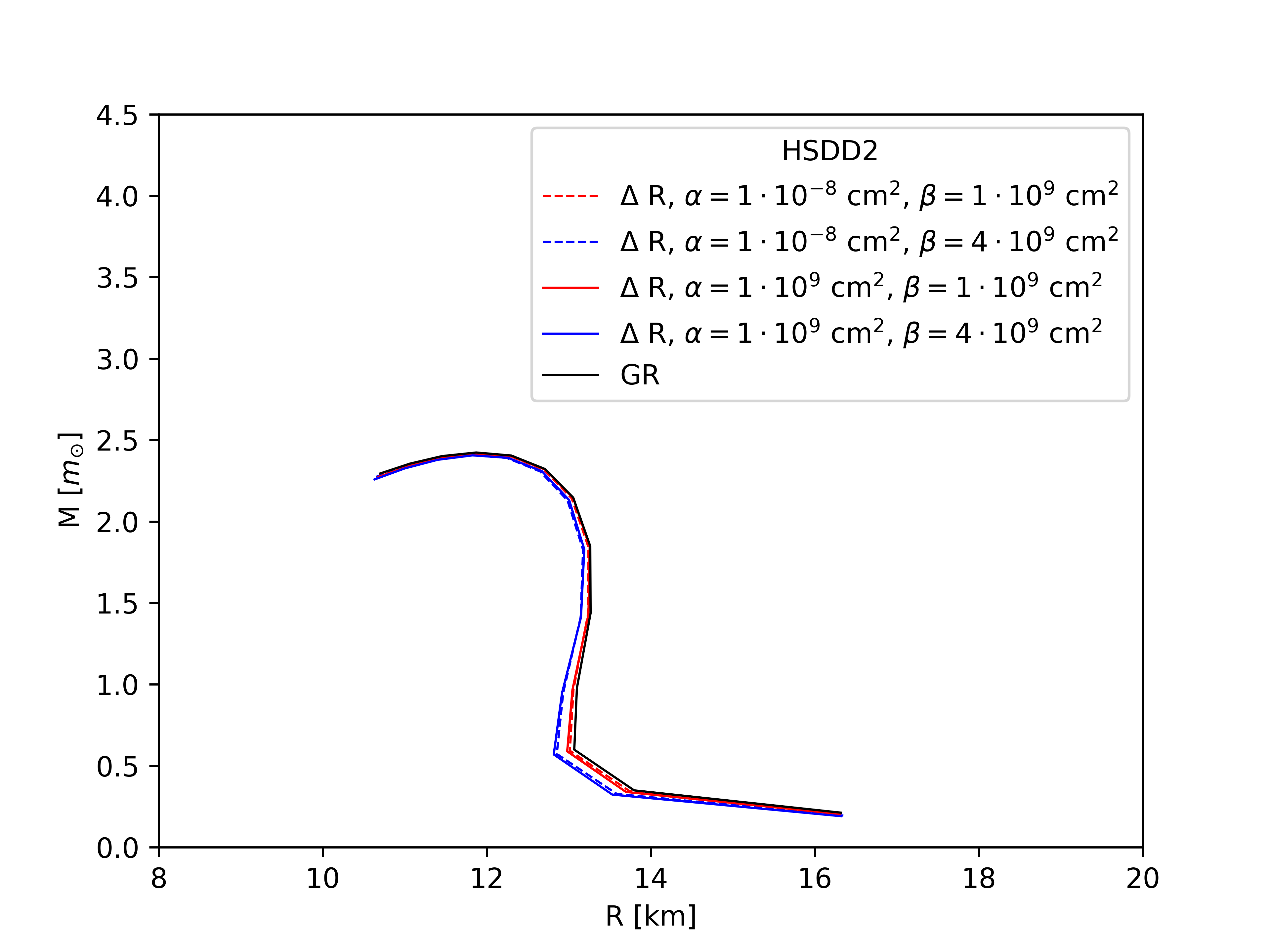}
\\
\centering
  \includegraphics[width=\linewidth]{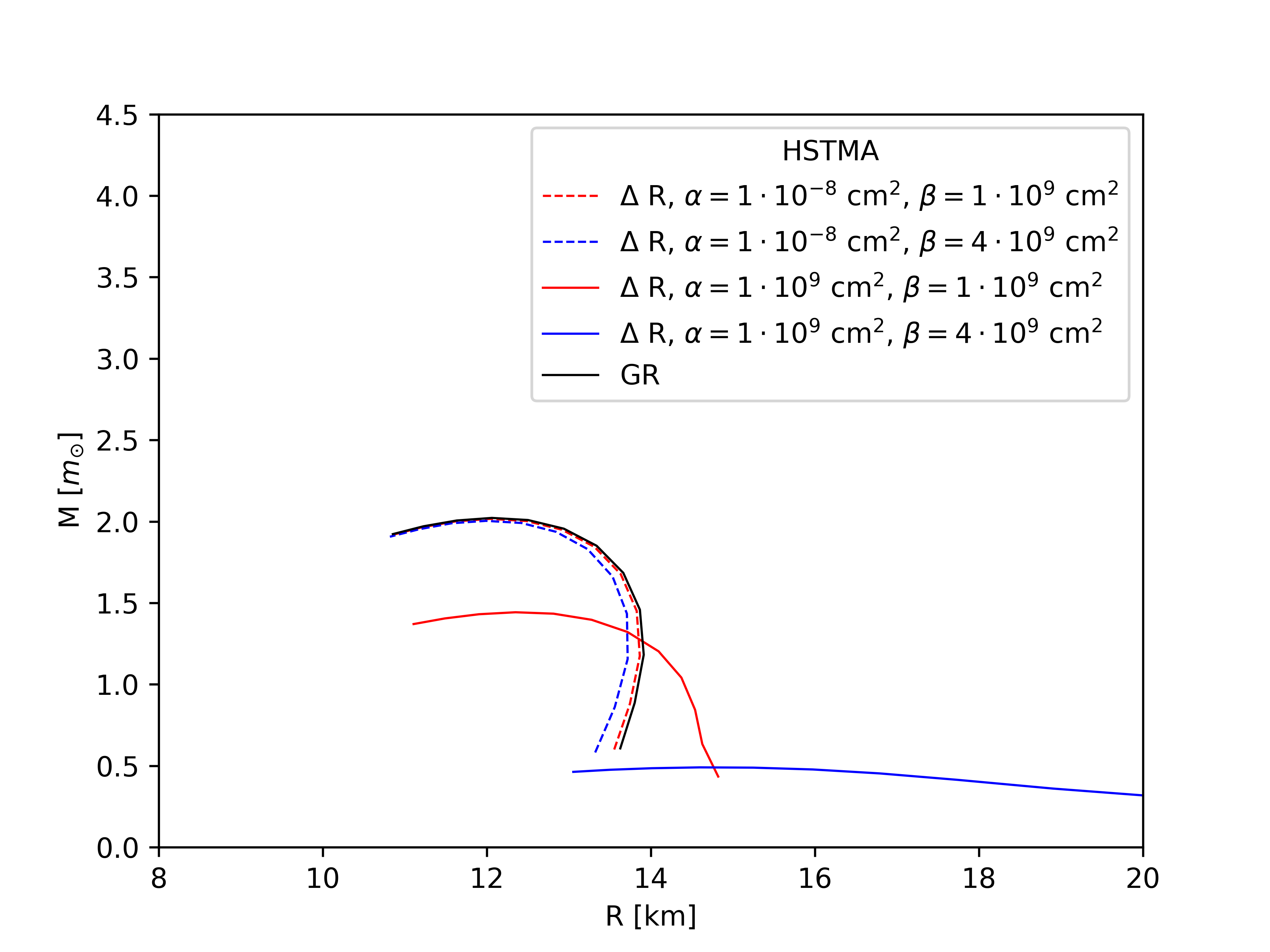}
\caption{Mass-radius relation for $f(R,Q)=R+\alpha R^2+\beta Q$ for HS(DD2) (upper panel) and HS(TDA) (lower panel) EoS.}
\label{fig:fRQ_tabulated}
\end{figure}

We show in Figs.~\ref{fig:fRQ_analytic} and \ref{fig:fRQ_tabulated} the mass-radius relation for two values of $\beta$, with and without the $R^2$ term (here the $R^2$ term is excluded by setting $\alpha$ to a very small value, for computational convenience). The effect of the $Q$ term appears to be generally small, the deviation from GR mostly coming from the $R^2$ term as we have seen in the previous section. The effect of the $Q$ term seems to be more pronounced for analytic EoS than for the tabulated ones, which may point to the problem with numerical derivatives that we discussed above. 

\begin{figure}[htbp]
  \centering
  \includegraphics[width=\linewidth]{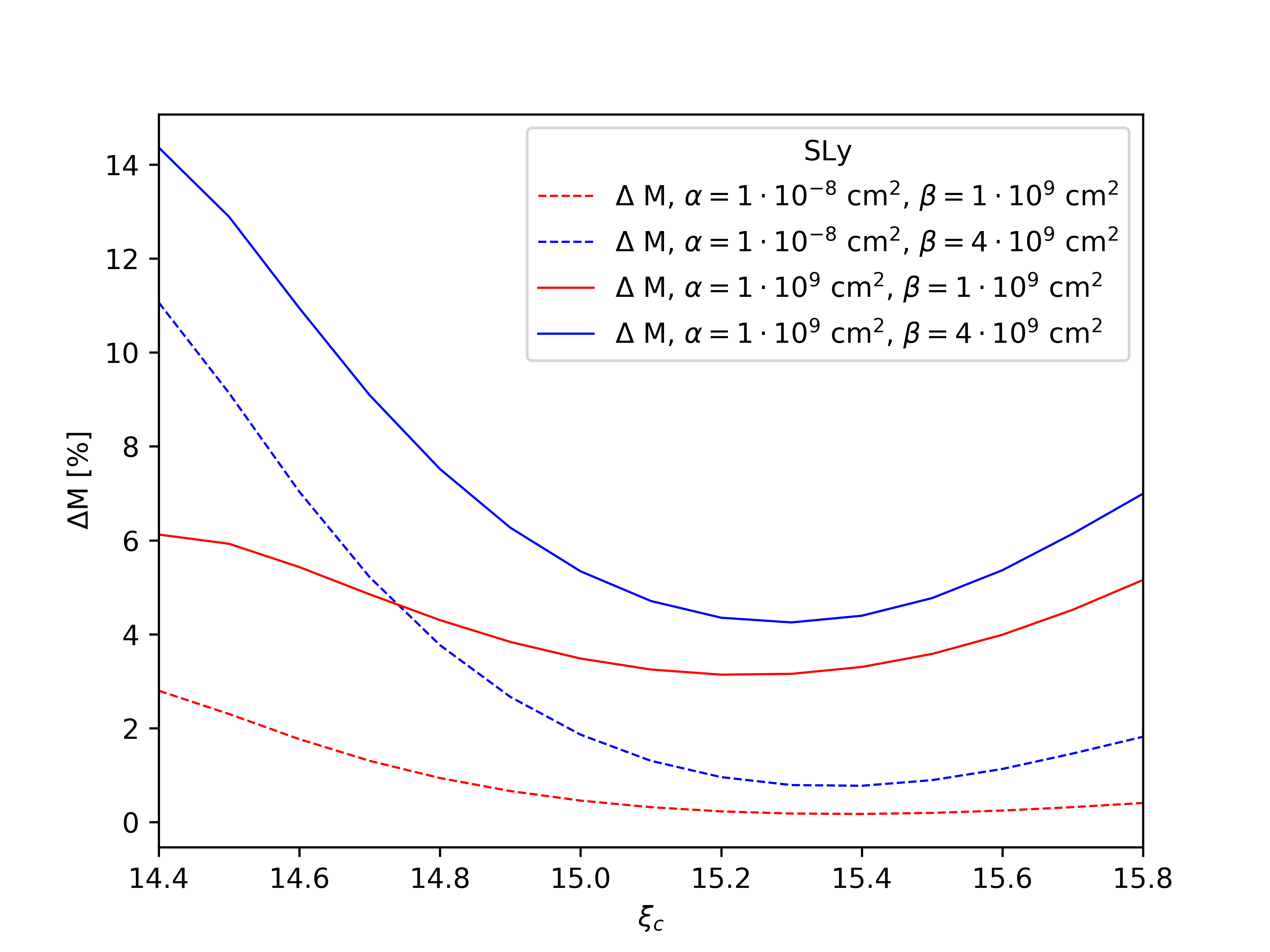}
\\
\centering
  \includegraphics[width=\linewidth]{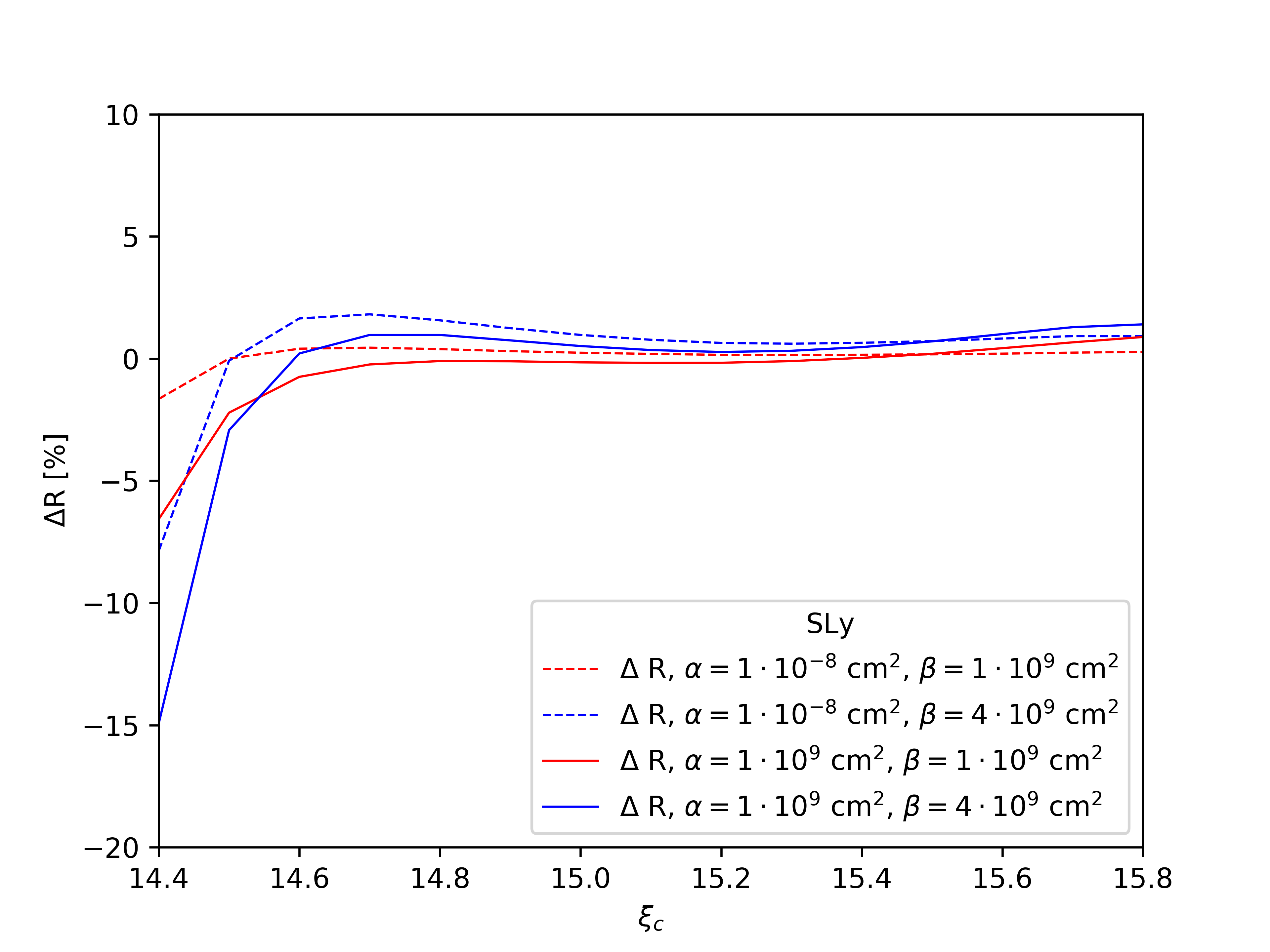}
\caption{Difference between the mass (upper panel) and the radius (lower panel) obtained, given a central density value $\xi$, with $f(R,Q)=R+\alpha R^2+\beta Q$ and GR, for the SLy EoS.}
\label{fig:DeltafRQ_SLY}
\end{figure}

\begin{figure}[htbp]
  \centering
  \includegraphics[width=\linewidth]{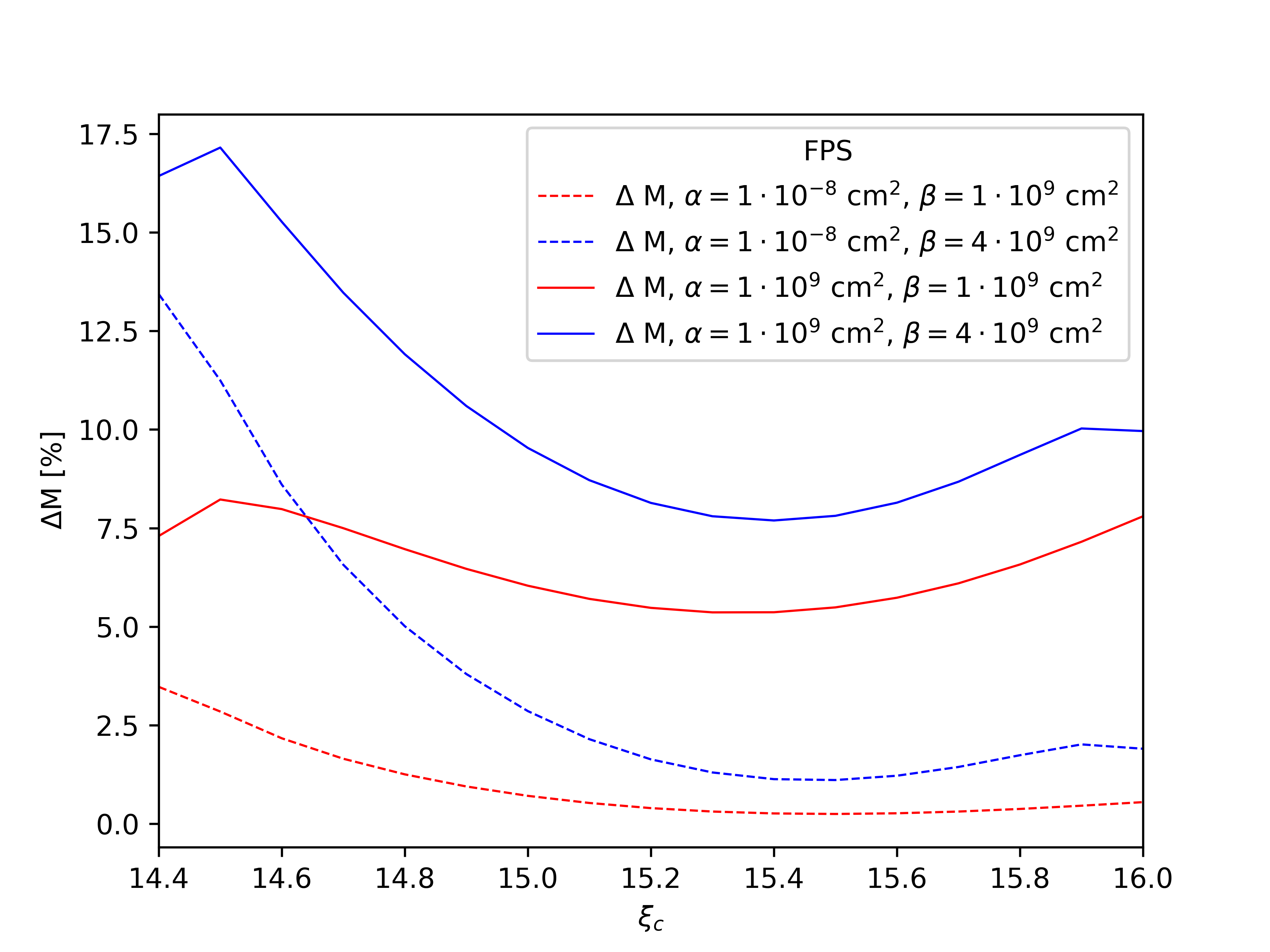}
\\
\centering
  \includegraphics[width=\linewidth]{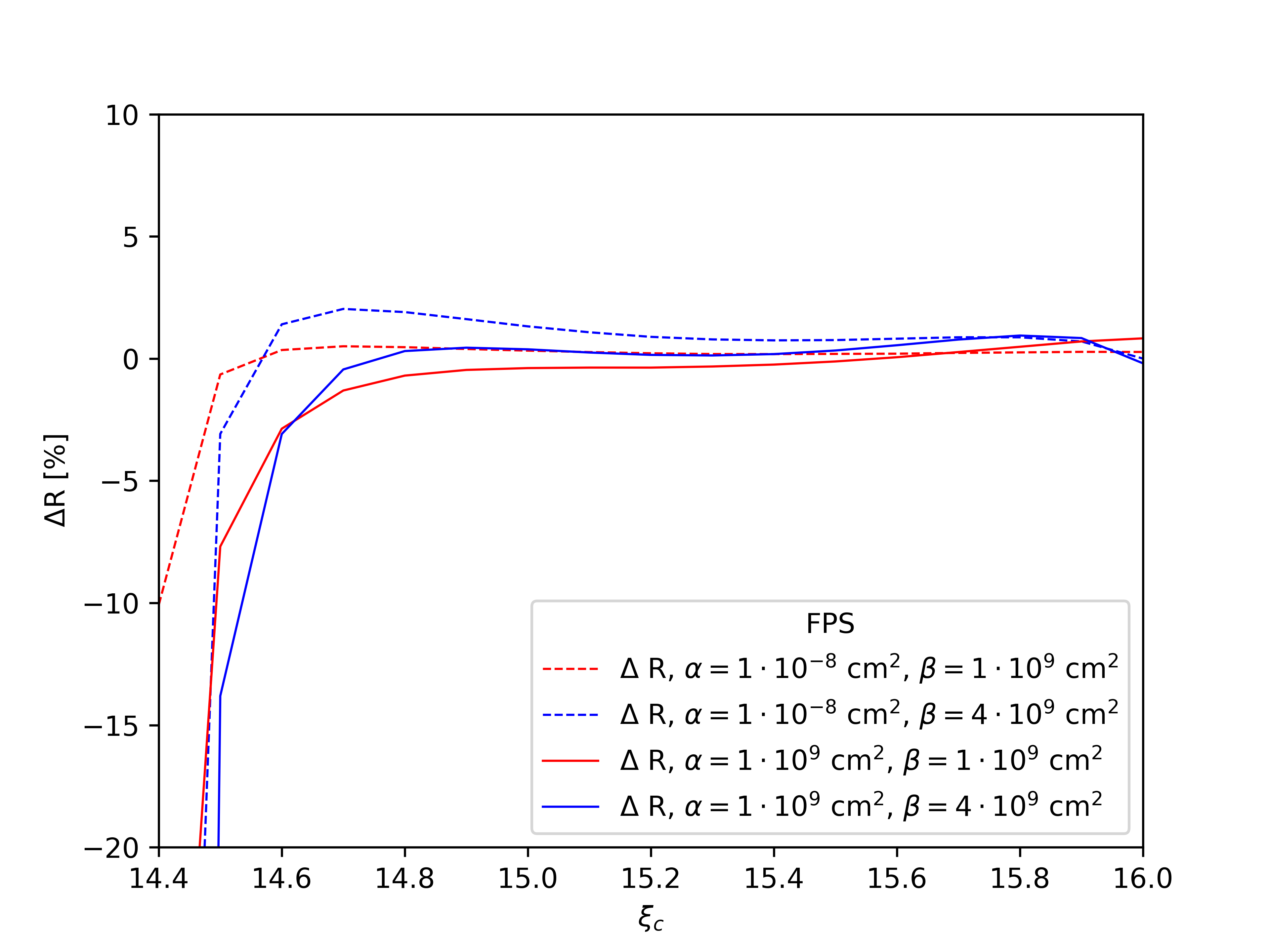}
\caption{Difference between the mass (upper panel) and the radius (lower panel) obtained, given a central density value $\xi$, with $f(R,Q)=R+\alpha R^2+\beta Q$ and GR, for the FPS EoS.}
\label{fig:DeltafRQ_FPS}
\end{figure}

\begin{figure}[htbp]
  \centering
  \includegraphics[width=\linewidth]{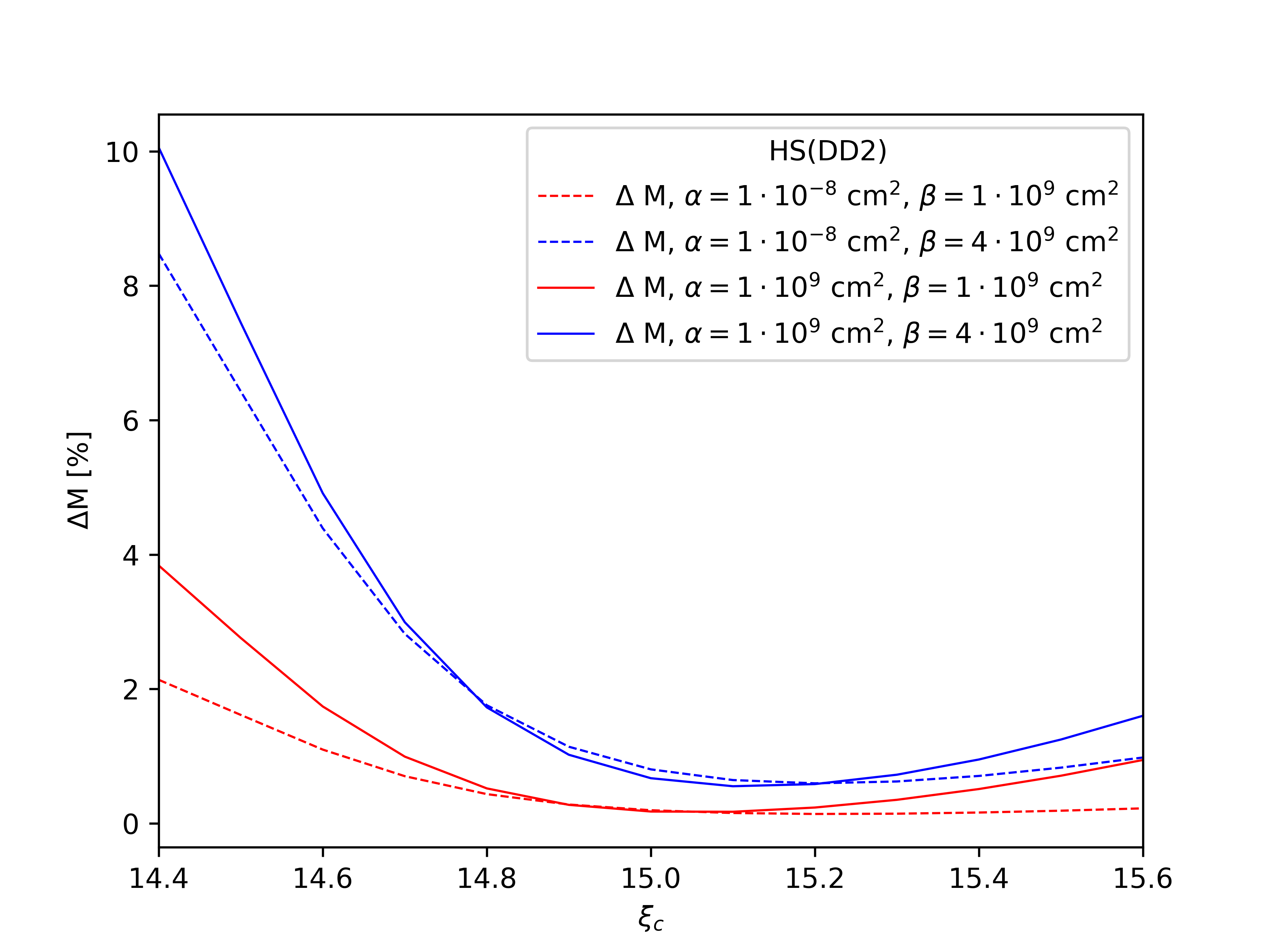}
\\
\centering
  \includegraphics[width=\linewidth]{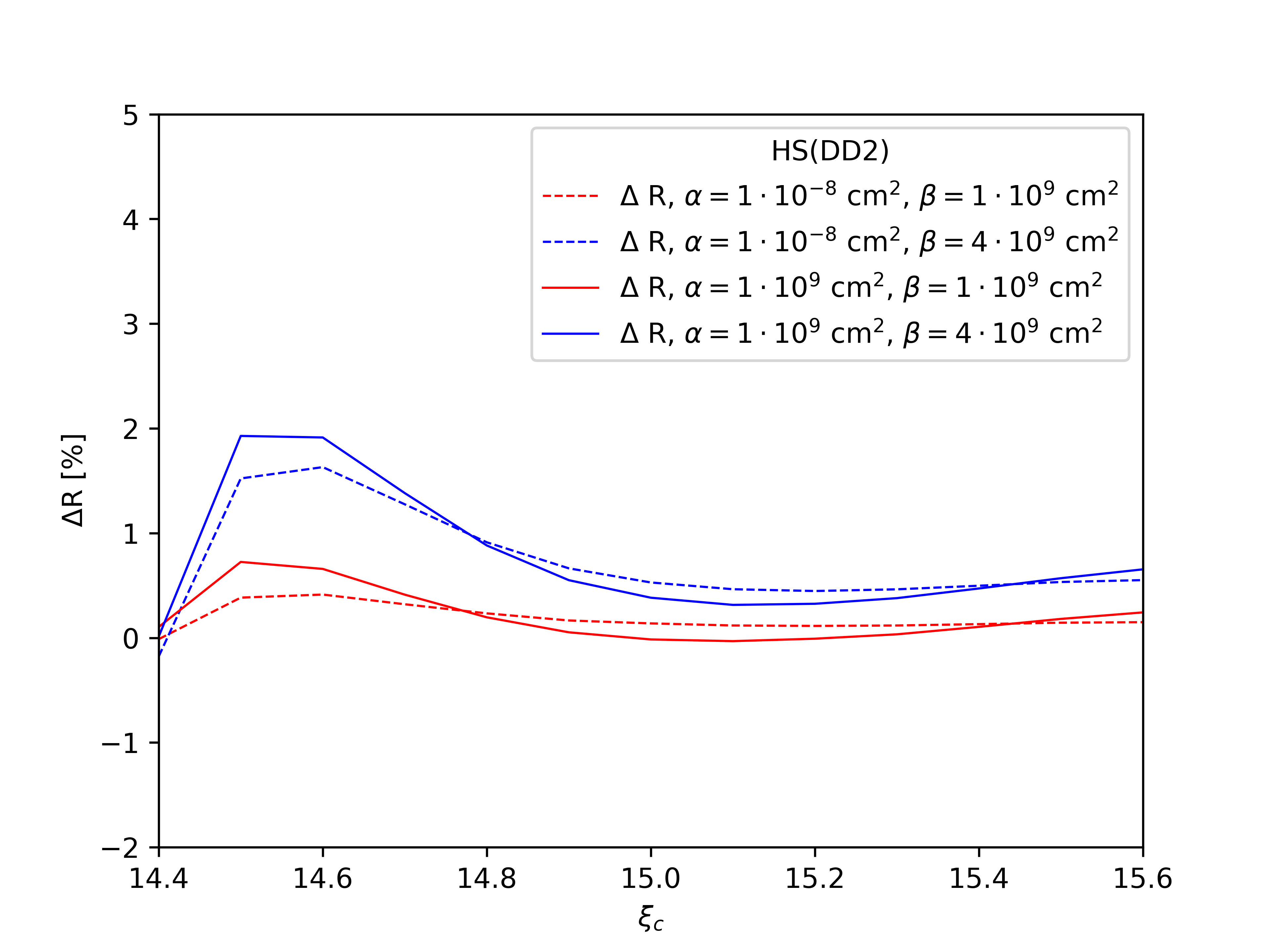}
\caption{Difference between the mass (upper panel) and the radius (lower panel) obtained, given a central density value $\xi$, with $f(R,Q)=R+\alpha R^2+\beta Q$ and GR, for the HS(DD2) EoS.}
\label{fig:DeltafRQ_HSDD2}
\end{figure}

\begin{figure}[htbp]
  \centering
  \includegraphics[width=\linewidth]{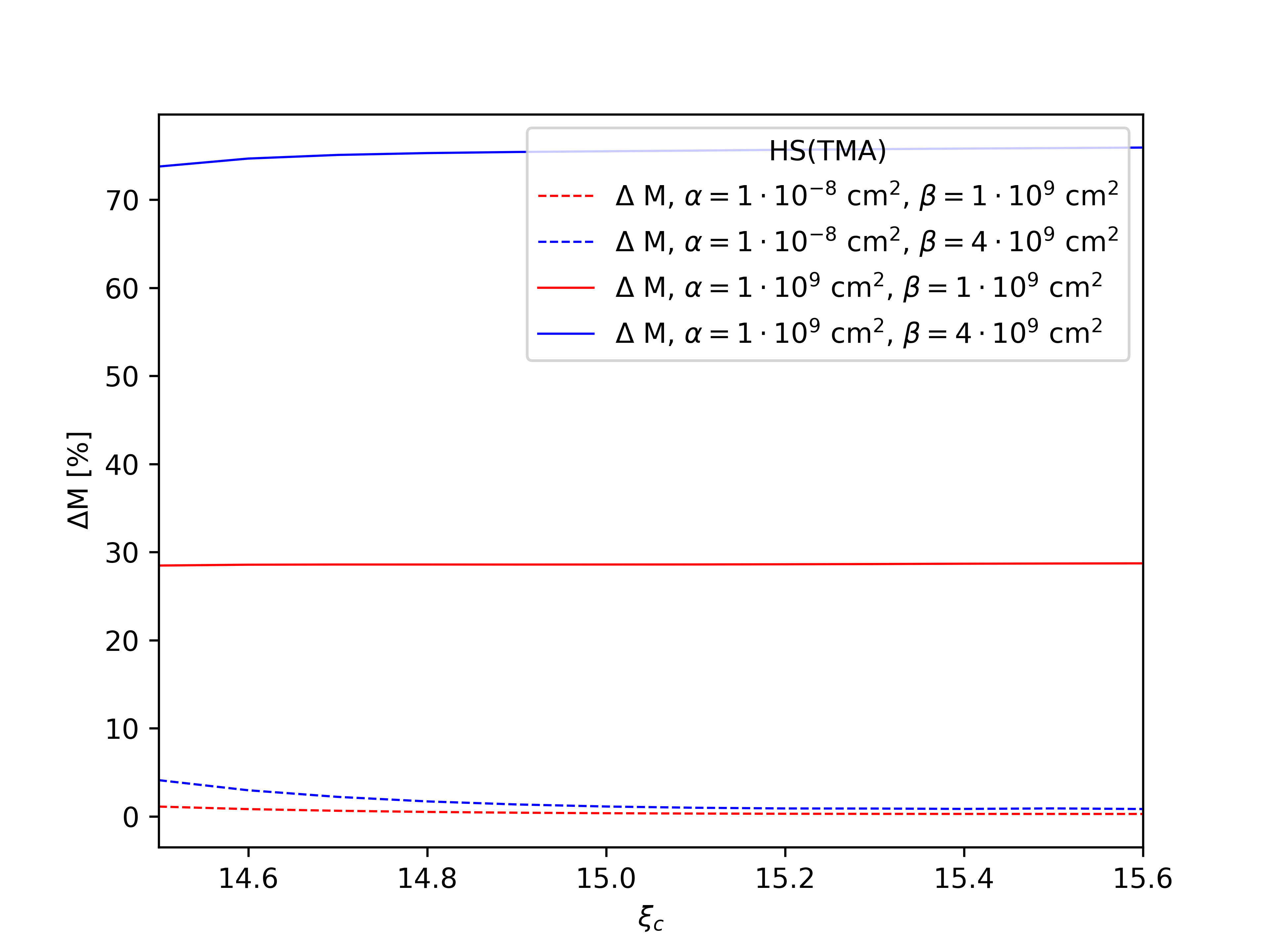}
\\
\centering
  \includegraphics[width=\linewidth]{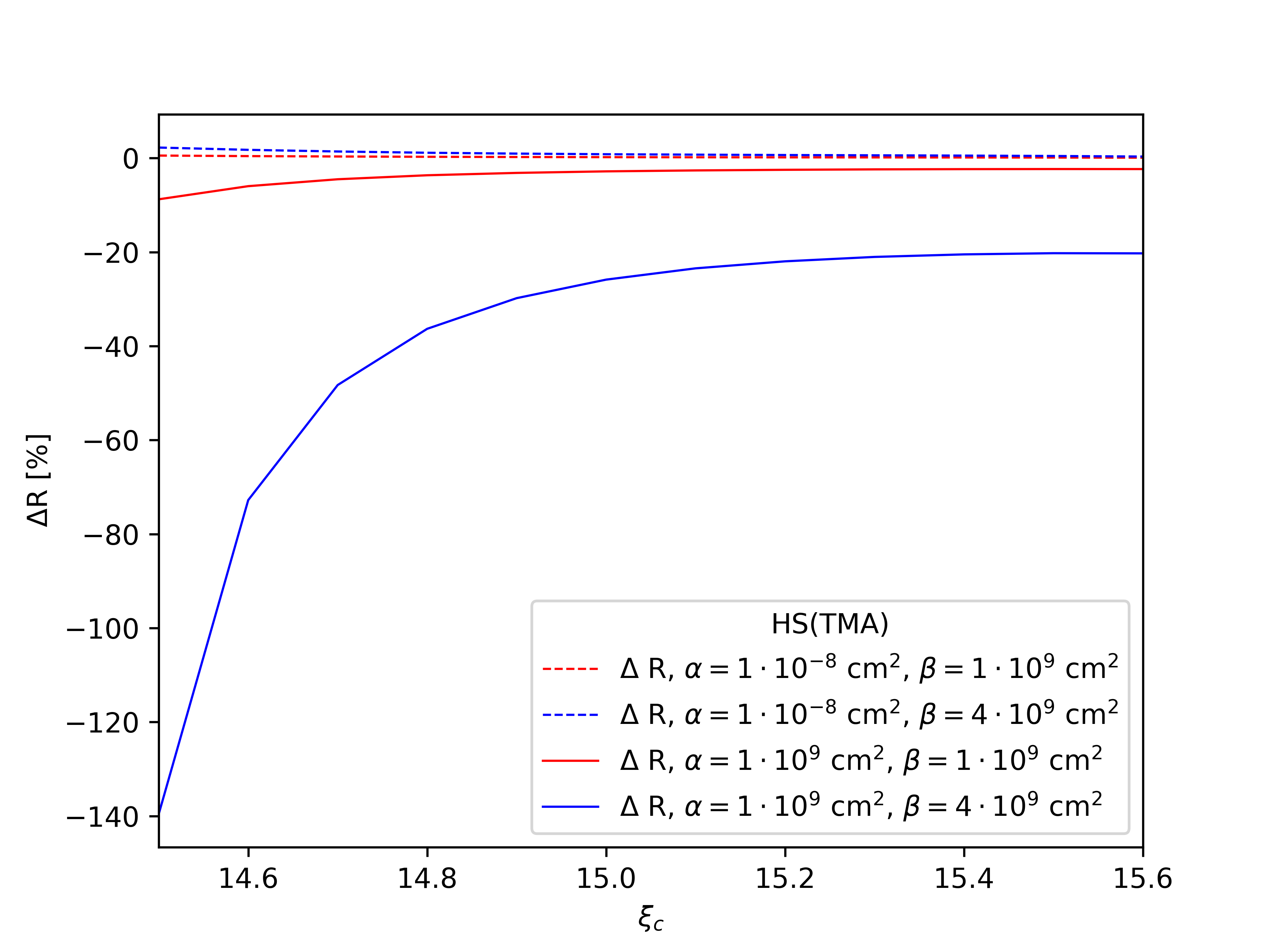}
\caption{Difference between the mass (upper panel) and the radius (lower panel) obtained, given a central density value $\xi$, with $f(R,Q)=R+\alpha R^2+\beta Q$ and GR, for the HS(TDA) EoS.}
\label{fig:DeltafRQ_HSTMA}
\end{figure}

As in the $f(R)$ case, it is interesting to plot separately the differences between the $f(R,Q)$ and the GR solutions, for the mass and the radius of a star. We show them in Figs.~\ref{fig:DeltafRQ_SLY}--\ref{fig:DeltafRQ_HSTMA}. In contrast to the $R^2$ term, for which the dependence of the mass and radius as a function of the central density showed a very different qualitative behaviour for different EoS, 
in the case of the $Q$ term these show a similar dependence on $\xi_c$ for all studied EoS. In all four cases the effect is somewhere below $5\%$ on the mass, with the exception of HS(DD2) which is up to $8\%$ for low central densities. 
%\comment{It is not true that the $Q$ term is less relavant for HS(TMA). The effect is similar if one zooms in on the plot. The huge deviation of the $R^2$ makes it harder to see since the yaxis is stretched. I made a zoom in plot to see it}. 
As for the $R^2$ case, it is interesting to note that the $Q$ terms influences the mass of the star much more than its radius, again with the exception of lowest values of the central density. In that region, the radius and mass effects compensate in a way such that they are not visible in the mass-radius plots.

\begin{figure}[htbp]
\centering
  \includegraphics[width=\linewidth]{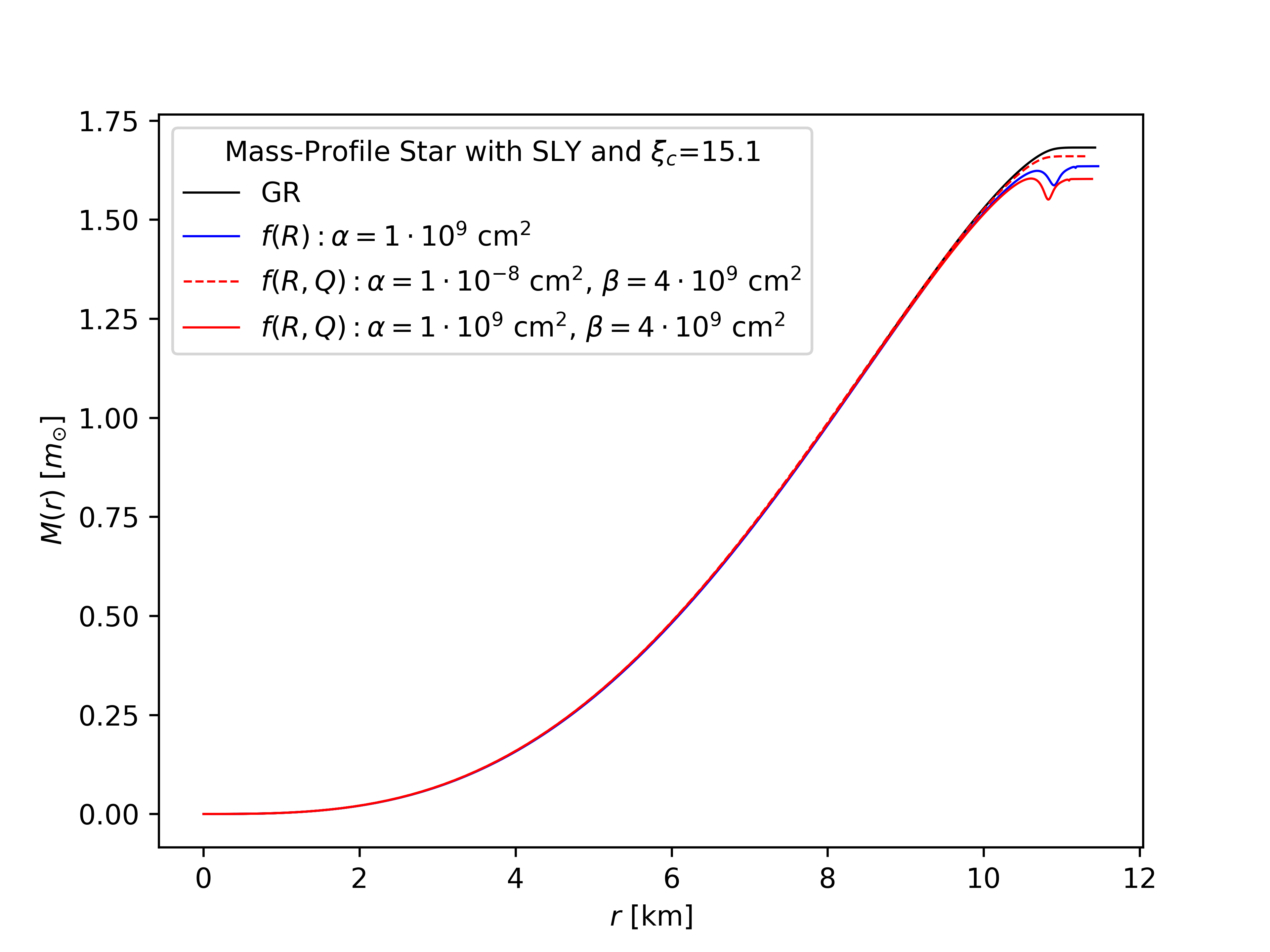}
  \caption{Metric function $M(r)$ as a function of the radial coordinate $r$, for a neutron star with a central density $\xi_c = 15.1$  and the SLy EoS.}
  \label{fig_m-r_profile}
\end{figure}

To conclude, we wish to mention a peculiar feature of the solutions we obtained for $f(R)$ gravities. In Fig.~\ref{fig_m-r_profile}, we show a star profile (that is, the metric function $M(r)$ as a function of the radial coordinate $r$) for the SLy EoS and $\xi=15.1$. The metric function $M(r)$ shows an unexpected cusp near the surface of the star. This feature, absent in GR, was already seen in \cite{Pannia:2016qbj} and predicted in \cite{Barausse:2007pn} (see also \cite{Pani:2012qd}), where it was argued that it invalidates Palatini $f(R)$ theories as physically viable. In \cite{Barausse:2007pn} it was also speculated that a $Q$ term may ameliorate this odd behaviour of the metric inside the star. As we see in Fig.~\ref{fig_m-r_profile}, it is true that an $f(Q)$ term (without $R^2$)  does not show a cusp\footnote{We have studied changes of $\beta$ by many orders of magnitude, and the cusp was absent in all cases.}. However, in an $f(R,Q)$ model (i.e. with both $R^2$ and $Q$ terms) the cusp remains; that is, the $Q$ term is not able to compensate for the unexpected behaviour generated by the $R^2$ term. 
Note that in \cite{Barausse:2007pn} the authors argue that for nonlinear $f(R)$ one always finds divergences at the surface of the star, which may in turn generate curvature divergences. For our numerical results, we have carefully investigated the appeareance of divergences in all terms of the equations of structure (e.g. divergences in $P_r$ or $P_{rr}$ which may result, upon integration, in finite values of $M(r)$) but found none. From our calculations, thus, we can only conclude that the cusp in Fig.~\ref{fig_m-r_profile} is simply a dynamical effect with no clear relation to the arguments in \cite{Barausse:2007pn}.
Note finally that, even though in Fig.~\ref{fig_m-r_profile} we showed  one particular solution only , in all cases we have studied in this paper the qualitative behaviour of the metric function $M(r)$ was analogous.

\section{Summary}\label{sec:summary}

In this paper we studied neutron stars solutions in Palatini $f(R)=R+\alpha R^2$ and $f(R,Q)=R+\alpha R^2+\beta Q$ gravities. We performed calculations using four different EoS and different values for $\alpha$ and $\beta$. 

Our study shows that the differences in the masses and radius of stars induced by modifications of the gravity Lagrangian can in fact be as large as the uncertainty induced by the use of different EoS. Notably, it is the mass of the star and not its radius that is mostly affected by modifications of GR. We have also discussed separately the effect of the $R^2$ and the $Q$ terms. The deviations from GR induced by the $R^2$ term are qualitatively different for different EoS. The $Q$ term instead shows a similar behaviour for all EoS, and in general smaller than the effect from the $R^2$ term.

Even though it was not a goal of our investigations, we found that some of the EoS which have previously been ruled out by neutron star observations, might become viable again if the $\alpha R^2$ correction is indeed part of the right theory of gravity. For negative $\alpha$ the maximum mass allowed by a given EoS tends to be increased. For some EoS, such as SLy or HS(TMA), it turns out that an $\alpha R^2$ term  can produce solutions as heavy as the MSP J0740+6620, the heaviest neutron star found so far \cite{Cromartie:2019kug}.

To further highlight the relevance of studies like the present one  it is interesting to mention the 2.6$M_{\odot}$ secondary component of GW190814 \cite{LIGOScientific:2020zkf}. Calculations using modified gravity theories indicate that it could be described as a neutron star (replacing MSP J0740+6620 as the heaviest known neutron star)
%or the lightest known black hole. 
Indeed, as shown in \cite{Astashenok:2020qds,Astashenok:2021peo} for metric $f(R)=R+\alpha R^2$ gravities, it is possible for certain EoSs to support neutron stars in the 2.6$M_{\odot}$ mass range. Interestingly, the authors showed that studying rotating stars also increases the maximum mass of neutron stars for those modified gravities. In this paper we observe a similar effect for Palatini $f(R)=R+\alpha R^2$ and $f(R,Q) = R+\alpha R^2 + \beta Q$, obtaining neutron star solutions in the 2.6$M_{\odot}$ mass range for the HS(TMA) EoS (see Fig.~\ref{fig_fR_hsdd2_hstma}). Alternatively, allowing a sufficiently negative value for $\alpha$ also leads to neutron stars in this mass range when using other EoS, like the analytic SLy EoS.

In this work we focused on the relatively simple problem of calculating the mass, radius and profiles of spherically symmetric stars. In the future, it is necessary to perform more detailed studies. For example, it is crucial to study the stability of a solution in order to understand whether it is viable as a physical solution. For example, it could be that the solutions beyond the GR mass limit are unstable and hence not realised in Nature. This is an even more pressing issue, considering the odd behaviour of the metric function $M(r)$ for $f(R)$ theories, as discussed in last section. Also, realistic neutron stars rotate and the system of equations solved herein provides only an estimate of their properties. Finally, on the technical side, it is necessary to establish reliable and unambiguous methods to calculate high-order derivatives of EoS, as we have seen.  

Clearly, no conclusive results can be obtained yet, at least until the ongoing efforts to nail down the EoS for a neutron star from first principles are successful. However, our study indicates  that, in the future, astrophysical observations may serve as discriminators of modified gravity. 

\begin{acknowledgements}
We thank G. Olmo for a critical reading of the manuscript.
\end{acknowledgements}

\bibliographystyle{spphys}       % APS-like style for physics
\bibliography{NStarsPalatini.bib}   % name your BibTeX data base

% \begin{thebibliography}{}
% Author, Article title, Journal, Volume, page numbers (year)
% Author, Book title, page numbers. Publisher, place (year)
% etc
% \end{thebibliography}

\end{document}